\documentclass[12pt]{article}

\usepackage{amsmath,amsfonts,amssymb,amsbsy}
\usepackage{graphicx,color}
\usepackage{graphics}

\usepackage{amsmath, amssymb,graphicx}
\usepackage{epsfig}
\usepackage{verbatim}
\usepackage{stmaryrd}
\usepackage{amsfonts}

\usepackage{epsf}

\newcommand{\be}{\begin{equation}}
\newcommand{\ee}{\end{equation}}
\newcommand{\ba}{\begin{eqnarray}}
\newcommand{\ea}{\end{eqnarray}}
\newcommand{\bea}{\begin{eqnarray}\begin{array}}
\newcommand{\eea}{\end{array}\end{eqnarray}}
\newcommand{\nn}{{\nonumber}}
\newcommand{\bc}{\begin{center}}
\newcommand{\ec}{\end{center}}

\newcommand{\beaa}{\begin{eqnarray}}
\newcommand{\eeaa}{\end{eqnarray}}

\newcommand{\bmx}{\begin{pmatrix}}
\newcommand{\emx}{\end{pmatrix}}

\newcommand{\mf}{\mathfrak}

\newcommand{\alf}{{\textstyle{\frac{1}{2}}}}
\newcommand{\half}{\frac{1}{2}}

\newcommand{\Smat}{{\mathcal S}}
\newcommand{\Refl}{{\mathcal R}}

\newcommand{\8}{{\infty}}

\newcommand{\eps}{\epsilon}
\newcommand{\tr}{\,{\rm tr}}

\newcommand{\bra}[1]{{\,\left<#1\right|}\,}
\newcommand{\ket}[1]{{\,\left|#1\right>}\,}

\newcommand{\up}{\uparrow}
\newcommand{\down}{\downarrow}

\newcommand{\calpha}{{\check\alpha}}
\newcommand{\cbeta}{{\check\beta}}
\newcommand{\cgamma}{{\check\gamma}}
\newcommand{\cdelta}{{\check\delta}}

\newcommand{\upd}{\upharpoonleft\hspace{-.1cm}\downharpoonright}

\newcommand{\twobh}{\boxslash\hspace{-.09cm}\boxslash}
\newcommand{\twobv}{\!\!\begin{array}{c}\boxslash\vspace{-.145cm}\end{array}\hspace{-.68cm}\begin{array}{c}\boxslash\vspace{.33cm}\end{array}\!\!}

\begin{document}



\begin{titlepage}
\begin{flushright}
{\bf \today} \\
DAMTP-08-64\\
DCPT-08/45\\
\end{flushright}
\begin{centering}
\vspace{.2in}

 {\Large {\bf Reflecting magnons from D7 and D5 branes}}

\vspace{.3in}

{\large D. H. Correa ${}^{a,1}$ and C. A. S. Young ${}^{b,2}$}\\
\vspace{.2 in}
${}^{a}${\emph{DAMTP, Centre for Mathematical Sciences \\
University of Cambridge\\ Wilberforce Road,
Cambridge CB3 0WA, UK}} \\
\vspace{.2in}
\vspace{.1 in}
${}^{b}${\emph{Department of Mathematical Sciences\\ University of Durham\\
South Road, Durham DH1 3LE, UK}}

%
\footnotetext[1]{{\tt D.Correa@damtp.cam.ac.uk,}\quad ${}^{2}${\tt charles.young@durham.ac.uk}}
\vspace{.5in}

{\bf Abstract}

\vspace{.1in}

\end{centering}

{We obtain the reflection matrices for the scattering of elementary magnons from certain open boundaries, corresponding to open strings ending on D7 and D5 branes in $AdS_5\times S^5$. In each case we consider two possible orientations for the vacuum state. We show that symmetry arguments are sufficient to determine the reflection matrices up to at most two unknown functions. The D7 reflection matrices obey the boundary Yang Baxter-Equation. This is automatic for one vacuum orientation, and requires a natural choice of ratio between two unknowns for the other.
In contrast, the D5 reflection matrices do not obey the boundary Yang Baxter-Equation. In both cases we show consistency with the existent weak and strong coupling results.}

\end{titlepage}

\tableofcontents
\section{Introduction}

In \cite{HMopen}, Hofman and Maldacena generalized the scattering theory of magnons in the planar limit of the AdS/CFT correspondence \cite{Bsu22}-\cite{AFZ} to include boundaries. The particular open boundary conditions they considered were those that arise when open strings end on certain D3-branes, known as giant gravitons.
On the gauge theory side, the  giant gravitons considered in \cite{HMopen} correspond to local operators involving the determinant of a given scalar field. The dual gauge theory description of these D3-branes and their open string excitations is thus entirely given in terms of pure $\mathcal N=4$ super Yang-Mills.
Subsequent work on the reflection of magnons in this context includes \cite{ChCo}-\cite{Palla}. Other interesting ways of introducing boundaries exist. In this paper, we shall consider the open boundary conditions associated to open strings ending on certain D5 and D7-branes \cite{KaRa,KaKa}, giving rise to dual gauge theories with less supersymmetry and with fundamental matter. We shall construct the all-loop reflection matrices for these cases, which will shed light on previous results (discussed below) concerning integrability at weak and strong coupling.

Let us begin by summarizing some details of the two 3+1-dimensional gauge theories we shall consider. Both are descendants of maximally supersymmetric Yang-Mills in which the usual $\mathcal N=4$ field multiplet, in the adjoint of the gauge group $SU(N)$, is supplemented by additional fields in the fundamental of $SU(N)$. Both are believed to be holographically dual to IIB string theory on $AdS_5\times S^5$ in the presence of certain additional probe branes \cite{DFO,KMMW}. With these gravity duals in mind, we shall speak of the D7 and D5 gauge theories. In this paper, we work solely in the strict planar limit $N\to \8$, in which these theories are conformal.

The D7 theory is $\mathcal N=2$ super-Yang Mills with a single chiral hypermultiplet of fundamental matter. Its gravity dual is IIB string theory on $AdS_5 \times S^5$ with a single\footnote{More generally, one can take a stack of $M\ll N$ D7 branes \cite{KMMW} but we shall focus for simplicity on the case with only $M=1$ flavour.} D7 brane which wraps the entire $AdS_5$ and a maximal $S^3$ of the $S^5$.

In the D5 case the dual theory has a single D5 brane which wraps a maximal $S^2$ of the $S^5$ and only an $AdS_4$ of the $AdS_5$. This $AdS_4\subset AdS_5$ defines a 2+1 dimensional \emph{defect} hypersurface of the 3+1 dimensional conformal boundary, which we take to be given by $x^3=0$. Since in the AdS/CFT dictionary fundamental matter in the gauge theory corresponds to open strings ending on the brane, in this theory the fundamental matter is constrained to live on the defect. For a single probe brane, this fundamental matter  consists of a single  3d hypermultiplet \cite{DFO}.

In either case, the addition of fundamental matter provides a new way to form local gauge-invariant operators. In addition to the usual closed chains of $\mathcal N=4$ fields, e.g.
\be \tr \,\,Z \dots\chi \dots  \phi  \dots Z \ee
constructed by taking the trace over the $SU(N)$ colour indices, there are also operators of the form
\be \bar q Z \dots  \chi \dots  \phi  \dots Zq \ee
where $q$, $\bar q$ are fields in, respectively, the fundamental and anti-fundamental of $SU(N)$. Of course, in the D5 case such operators exist only on the defect. Following the original insight of \cite{MZ}, in the planar limit $N\to\8$ such operators can be thought of as open spin chains, with the dilatation operator playing the role of the Hamiltonian.

Just as in the case of the closed chains, \cite{BS}
at first order in the 't Hooft coupling $\lambda$ the Hamiltonian has only nearest-neighbour interactions.  Indeed, away from the endpoints of the chain the Hamiltonian is the same as in the unperturbed $\mathcal N=4$ case, because all corrections from processes involving virtual fundamental fields are suppressed by factors of $1/N$. It is only near the ends of the chain that the perturbation of $\mathcal N=4$ has any effect at all. This will be an important observation for us in what follows.

An important question is whether the boundary conditions are integrable or not. Working up to first order in the 't Hooft coupling $\lambda$ it was shown in \cite{DeMa,ErMa} that both the D5 and D7 cases yield integrable boundary conditions, at least in the $SO(6)$ sector (further work about integrability in dCFT includes \cite{McSw}-\cite{OTY}). That is, the one-loop spin-chain Hamiltonian is one of a family of conserved quantities encoded in a transfer matrix, which in turn is built out of a bulk Lax matrix $L$ together with a boundary reflection matrix $K$, following the usual techniques of integrable open spin chains \cite{Skly}. (Further results concerning integrable open spin chains in the AdS/CFT context can be found in \cite{ChenWangWu}.) On the other hand, in the opposite regime $\lambda\to\8$, open strings ending on D7-branes were shown to be classically integrable by explicit construction of the corresponding non-local conserved charges \cite{Mann}. However, the same technique failed for open strings ending on D5-branes \cite{Mann}.

At weak coupling beyond 1-loop, the complications familiar from the closed-chain operators \cite{B2003} will arise: in particular, the interactions become long-range, and the length of the chain will not remain constant. To make progress one is lead, following \cite{BMN}, to consider operators consisting almost entirely of a single scalar field $Z$, which is regarded as the vacuum state, with a few other fields -- ``magnons'' -- scattered along the chain. More precisely, one chooses a preferred $R$-charge $J$ and considers states in which both $J$ and the classical dimension $\Delta$ are large, but with $\Delta-J$ held finite. The vacuum state $Z$ has $\Delta-J=1-1=0$ and the elementary magnons are those fields with $\Delta - J=1$. In this way it is possible to ignore the microscopic details of the spin chain and focus on the macroscopic scattering theory \cite{AFS,Staudacher} of the magnons. Symmetry considerations alone turn out to be powerful enough to determine the two-particle S-matrix up to a single overall factor \cite{Bsu22}.

As we noted above, deep in the ``bulk'' of an open spin chain the theories we consider are indistinguishable from pure $\mathcal N=4$, so these symmetry arguments still apply and the bulk S-matrix is unmodified. What remains to be determined is the scattering behaviour of magnons off the end of the chain. Note that the relative orientation between the preferred $R$-charge of vacuum and the spherical factor of the worldvolume of the
brane will affect the symmetries preserved by the reflection. We shall discuss a couple of inequivalent possibilities for both the D5 and the D7 case. As we shall see, in certain cases the boundary itself can have an excitation attached to it.

The full actions for the D7 and D5 gauge theories can be found in \cite{ErMa,DeMa}. For our purpose -- that of constructing reflection matrices and determining whether they are integrable -- it will suffice, just as in the case of strings ending on maximal giant gravitons \cite{HMopen}, to perform in each case a careful analysis of the symmetries preserved by the boundary and the representation content of the theory with respect to these symmetries. We do this, and make certain checks against one-loop results, for the D7 case in section \ref{sec:D7} below, before turning to the D5 case in section \ref{sec:D5}. Some conclusions are given in section \ref{conc}.

\vspace{1cm}
\textbf{Note added in May 2011:} Some time after the publication of this article we discovered a sign error in the boundary Yang-Baxter equations of section 4.3. Once this error is corrected, the D5-brane reflection matrix, as presented, \emph{does} obey the boundary Yang-Baxter equation, for a certain choice of the free ratio. The crucial error comes from omitting a graded permutation when left and right factors of the bulk magnon are exchanged; a complete discussion with the correct boundary Yang-Baxter equation can now be found in \cite{CRY}.

\section{Symmetries}
We begin by recalling the definition of the superalgebra $psl(4|4)$ of superconformal symmetries of $\mathcal N=4$ SYM, following conventions similar to those of \cite{BeisertThesis}. The generators of the even subalgebra include $\mathfrak D, \mathfrak{L}^\alpha_{~\beta}, \tilde{\mathfrak{L}}^{\dot \alpha}_{~\dot\beta},
 {\mathfrak{R}}^{\rm a}_{~{\rm b}}$
of, respectively, dilatations, the $sl(2)\times sl(2)$ of Lorentz and the $sl(4)=so(6)$ of R-symmetry. Here  \be \alpha,\beta,\ldots=+,-,\quad \dot\alpha,\dot\beta,\ldots=\dot+,\dot-,\quad {\rm a},{\rm b},\ldots=1,2,3,4.\ee
The remaining generators are the translations $\mathfrak{P}^{\dot\alpha\beta}$, conformal transformations $\mathfrak{K}_{\alpha\dot\beta}$, supersymmetries $(\mathfrak{Q}^{\alpha}_{~\rm a}$, $\tilde{\mathfrak{Q}}^{\dot\alpha{\rm a}})$, and superconformal transformations $(\mathfrak{S}^{\rm a}_{~\alpha}, \tilde{\mathfrak{S}}_{{\rm a}\dot\alpha})$. Their dimensions are $1,-1,\half,-\half$ and they transform canonically according to the indices they carry:
\begin{align}
[\mathfrak{L}^\alpha_{~\beta},\mathfrak{J}^\gamma]
&=\delta^{\gamma}_\beta \mathfrak{J}^\alpha-\tfrac12\delta^{\alpha}_\beta \mathfrak{J}^\gamma\,,
\qquad
[\mathfrak{L}^\alpha_{~\beta},\mathfrak{J}_{\gamma}]
= -\delta_{\gamma}^\alpha \mathfrak{J}_\beta+\tfrac12\delta^{\alpha}_\beta \mathfrak{J}_{\gamma} \,,
\label{eq1}\nn
\\
[\tilde{\mathfrak{L}}^{\dot\alpha}_{~\dot\beta}, \mathfrak{J}^{\dot\gamma}]
&= \delta^{\dot\gamma}_{\dot\beta} \mathfrak{J}^{\dot\alpha}-\tfrac12\delta^{\dot\alpha}_{\dot\beta}
\mathfrak{J}^{\dot\gamma} \,,
\qquad
[\tilde{\mathfrak{L}}^{\dot\alpha}_{~\dot\beta},\mathfrak{J}_{\dot\gamma}]
= -\delta_{\dot\gamma}^{\dot\alpha} \mathfrak{J}_{\dot\beta}+\tfrac12 \delta^{\dot\alpha}_{\dot\beta}
 \mathfrak{J}_{\dot\gamma} \,,\nn
 \\
[ {\mathfrak{R}}^{\rm a}_{~{\rm b}}, \mathfrak{J}^{\rm c}]
&=
\delta^{\rm c}_{\rm b} \mathfrak{J}^{\rm a}-\tfrac14 \delta^{\rm a}_{\rm b} \mathfrak{J}^{\rm c}\,,
\qquad\;
[ {\mathfrak{R}}^{\rm a}_{~{\rm b}}, \mathfrak{J}_{\rm c}]
=-\delta_{\rm c}^{\rm a} \mathfrak{J}_{\rm b}+\tfrac14 \delta^{\rm a}_{\rm b} \mathfrak{J}_{\rm c}\,.
\end{align}
The remaining non-trivial commutation relations are
\begin{align}
[\mathfrak{K}_{\alpha\dot\beta},\mathfrak{P}^{\dot\gamma\delta}]
&=  \delta^{\dot\gamma}_{\dot\beta} \mathfrak{L}^\delta_{~\alpha}
+  \delta_{\alpha}^{\delta} \tilde{\mathfrak{L}}^{\dot\gamma}_{~\dot\beta}
+ \delta^{\dot\gamma}_{\dot\beta} \delta_{\alpha}^{\delta} \mathfrak{D}\,,\nn
\\
[\mathfrak{S}^{\rm a}_{~\alpha},\mathfrak{P}^{\dot\beta\gamma}]
&=\delta^{\gamma}_\alpha \tilde{\mathfrak{Q}}^{\dot\beta{\rm a}}\,,
\qquad \;\;
[{\mathfrak{Q}}^{\gamma}_{~{\rm a}},\mathfrak{K}_{\alpha\dot\beta}]
= -\delta^{\gamma}_{\alpha}\tilde{\mathfrak{S}}_{{\rm a}\dot\beta} \,,\nn
\\
[\tilde{\mathfrak{S}}_{{\rm a}\dot\alpha},\mathfrak{P}^{\dot\beta\gamma}]
&=\delta_{\dot\alpha}^{\dot\beta} {\mathfrak{Q}}^{\gamma}_{~{\rm a}}\,,
\qquad \;\;
[\tilde{\mathfrak{Q}}^{\dot\gamma{\rm a}},\mathfrak{K}_{\alpha\dot\beta}]
= -\delta^{\dot\gamma}_{\dot\beta}\mathfrak{S}^{\rm a}_{~\alpha} \,,\nn
 \\
\{\tilde{\mathfrak{S}}_{{\rm a}\dot\alpha},\mathfrak{S}^{\rm b}_{~\beta}\}
&=\delta_{\rm a}^{\rm b} {\mathfrak{K}}_{\beta\dot\alpha}\,,
\qquad
\{\tilde{\mathfrak{Q}}^{\dot\alpha{\rm a}},\mathfrak{Q}^{\beta}_{~\rm b}\}
= \delta^{\rm a}_{\rm b} \mathfrak{P}^{\dot\alpha\beta} \,,\nn
\\
\{\mathfrak{S}^{\rm a}_{~\alpha},\mathfrak{Q}^{\beta}_{~\rm b}\}
&= \delta^{\rm a}_{\rm b} \mathfrak{L}^\beta_{~\alpha}
+  \delta^\beta_{\alpha}  {\mathfrak{R}}^{\rm a}_{~\rm b}
+\tfrac12 \delta^{\rm a}_{\rm b} \delta^\beta_{\alpha} \mathfrak{D} \,,\nn
\\
\{\tilde{\mathfrak{S}}_{{\rm a}\dot\alpha},\tilde{\mathfrak{Q}}^{\dot\beta\rm b}\}
&=\delta_{\rm a}^{\rm b}\tilde{\mathfrak{L}}^{\dot\beta}_{~\dot\alpha}
-  \delta^{\dot\beta}_{\dot\alpha}  {\mathfrak{R}}^{\rm b}_{~\rm a}
+\tfrac12 \delta^{\rm b}_{\rm a} \delta^{\dot\beta}_{\dot\alpha} \mathfrak{D}\, .
\end{align}

In this language the fields appearing in the $\mathcal N=4$ action are the gauge connection
$A^{\alpha \dot \alpha}$, the fermions $\Psi^{\alpha \rm b}$, $\tilde \Psi^{\dot\alpha}_{\rm b}$ and the scalars $\phi^{\rm [ab]}$.

At the boundary of the scattering theory we consider -- that is, near the endpoints of an operator of the form
\be \bar q Z  \dots Z\chi Z \dots Z\phi Z \dots Z q, \ee
$psl(4|4)$ symmetry is broken in two ways:  by the choice of Bethe vacuum state $Z$, and explicitly by the extra terms involving fundamental matter added to the original $\mathcal N=4$ action. Equivalently, on the gravity side $psl(4|4)$ is broken by our choice of angular momentum generator $J\in so(6)$, and by the presence of the D-brane.

\section{D7-brane}\label{sec:D7}
\subsection{Boundary symmetries}
Consider a D7-brane whose worldvolume wraps $AdS_5$ entirely and the maximal
$S^3\subset S^5$ defined, without loss of generality, by $X_5=X_6=0$. The $so(6)$ symmetry is broken to
$so(4)_{1234}\times so(2)_{56}$. In our conventions (given in appendix \ref{conventions}) the generators of the $so(4)\cong sl(2)\times \widetilde{sl}(2)$ are then $R^{a}_{~b}$ and $\tilde{R}^{\dot a}_{~\dot b}$, with
\begin{eqnarray}
&R^{{1}}_{~ {2}} =  {\mathfrak{R}}^{1}_{~2}\,,
\qquad  \qquad   \qquad   \qquad\;\;\;
&\tilde{R}^{\dot1}_{~\dot2} =  - {\mathfrak{R}}^{4}_{~3}\,,\nn
\\
& R^{2}_{~1} =  {\mathfrak{R}}^{2}_{~1} \,,
\qquad \qquad  \qquad \qquad\;\;\;
& \tilde{R}^{\dot2}_{~\dot1} =   - {\mathfrak{R}}^{3}_{~4}\,,\nn
\\
&  R^{1}_{~1} = - R^2{}_2 = \alf \mf R^1{}_1 - \alf \mf R^2{}_2\,,
\quad
& \tilde{R}^{\dot1}_{~\dot1} =  - \tilde R^{\dot 2}{}_{\dot 2} = \alf \mf R^4{}_4 - \alf \mf R^3{}_3\,,
\end{eqnarray}
and the supersymmetries with indices $\rm a = 3,4$ become\footnote{In what follows, this naming of indices will ensure that both copies of $psl(2|2)$ have the standard anti-commutation relations: \begin{align}
\{\mathfrak{S}^{a}_{~\alpha},\mathfrak{Q}^{\beta}_{~ b}\}
&= \delta^{ a}_{ b} \mathfrak{L}^\beta_{~\alpha}
+  \delta^\beta_{\alpha} {{R}}^{a}_{~b}
+\tfrac12 \delta^{a}_{b} \delta^\beta_{\alpha} (\mathfrak{D}-J_{56}) \,,\nn
\\
\{\tilde{\mathfrak{S}}^{\dot a}_{~\dot\alpha},\tilde{\mathfrak{Q}}^{\dot\beta}_{~ \dot b}\}
&=\delta^{\dot a}_{\dot b}\tilde{\mathfrak{L}}^{\dot\beta}_{~\dot\alpha}
+\delta^{\dot\beta}_{\dot\alpha} {\tilde{R}}^{\dot a}_{~\dot b}
+\tfrac12 \delta^{\dot a}_{\dot b} \delta^{\dot\beta}_{\dot\alpha} (\mathfrak{D}-J_{56})\, .\nn
\end{align}}
\be\mf Q^\alpha{}_{\dot 1} = - \mf Q^\alpha{}_{4},\quad \mf Q^\alpha{}_{\dot 2} = \mf Q^\alpha{}_3,
 \quad \tilde{\mf Q}^{\dot\alpha}{}_{\dot 1} =  \tilde{\mf Q}^{\dot\alpha 3},\quad
      \tilde{\mf Q}^{\dot\alpha}{}_{\dot 2} = \tilde{\mf Q}^{\dot\alpha 4}\,, \ee
\be\mf S^{\dot 1}{}_\alpha = \mf S^{4}{}_\alpha, \quad \mf S^{\dot 2}{}_\alpha = -\mf S^3{}_\alpha,
\quad \tilde{\mf S}^{\dot 1}{}_{\dot \alpha} = \tilde{\mf S}_{3\dot \alpha}, \quad
      \tilde{\mf S}^{\dot 2}{}_{\dot \alpha} = \tilde{\mf S}_{4\dot \alpha}\,.  \ee
The D7-brane preserves the half of the background supersymmetries that are right-handed with respect to this $so(4)$ \cite{KMMW}\footnote{This is easy to see if one regards the both the stack of $N$ $D3$'s and the $D7$ as probe branes in flat space}, that is, those carrying dotted latin indices $\dot a,\dot b, \dots$:
\be
\mf Q^\alpha{}_{\dot a}, \quad \tilde{\mf Q}^{\dot \alpha}{}_{\dot a},
 \quad \mf S^{\dot a}{}_\alpha, \quad \tilde{\mf S}^{\dot a}{}_{\dot \alpha}\,.
\label{D7susys}
\ee
Some of these symmetries will be further broken by the vacuum state.
The resulting residual symmetry will depend on how this vacuum is chosen.
Next we consider two possibilities.

~

\noindent
{\bf Bulk vacuum state $Z$.} \label{d7Z} We take the preferred R-charge $J\in so(6)$ to be
\be J_{56} =- \alf\mf R^1{}_1 - \alf\mf R^2{}_2 + \alf\mf R^3{}_3 + \alf\mf R^4{}_4\,, \ee
which rotates the directions transverse to the brane and preserves the full $sl(2)\times \widetilde {sl}(2)$ symmetry. The corresponding spin-chain vacuum is
\be Z= X_5 + i X_6 = \phi^{34}. \ee
The 16 supersymmetries neutral under $\mf D-J_{56}$, and so preserving $Z$, are as usual
\be
\mf Q^\alpha{}_{a}, \quad \tilde{\mf Q}^{\dot \alpha}{}_{\dot a},
 \quad \mf S^{a}{}_\alpha, \quad \tilde{\mf S}^{\dot a}{}_{\dot \alpha}
\ee
but of these only (cf \ref{D7susys})
\be
\tilde{\mf Q}^{\dot \alpha}{}_{\dot a}, \quad \tilde{\mf S}^{\dot a}{}_{\dot \alpha}
\ee
are supersymmetries of the D7. Thus, of the $psl(2|2)\otimes\widetilde{psl}(2|2)\ltimes \mathbb R^3$ symmetry algebra of the scattering theory in the bulk \cite{Bsu22}, the residual symmetry algebra at the boundary for this choice of vacuum is
\be
sl(2)_{\mf L} \times sl(2)_{R} \times \widetilde{psl}(2|2)_{\tilde{\mf L}, \tilde R, \tilde{\mf Q} , \tilde{\mf S}} \ltimes \mathbb R^3.
\label{D7zsim}
\ee

\noindent
{\bf Bulk vacuum state $X$.} It will also be useful work with the vacuum
\be X = X_1 + i X_2 = \phi^{14}. \ee
Then $J$ is
\be J_{12}
   = \alf \mf R^1{}_1 - \alf \mf R^2{}_2 - \alf \mf R^3{}_3 + \alf \mf R^4{}_4
   =  R^1{}_1 +  \tilde R^{\dot 1}{}_{\dot 1} \ee
Of the supersymmetries (\ref{D7susys}) of the D7-brane, those neutral under $\mf D - J_{12}$ and so preserving $X$ are
\be\mf Q^\alpha{}_{3}, \quad \tilde{\mf Q}^{\dot \alpha 4},
\quad \mf S^{3}{}_\alpha, \quad \tilde{\mf S}_{4\dot\alpha}
\label{D7Xsusys}\ee
The $sl(2)_L \times sl(2)_R$ is broken to the $u(1)$ of
\be J_{34} = -\alf \mf R^1{}_1 + \alf \mf R^2{}_2 - \alf \mf R^3{}_3 + \alf \mf R^4{}_4
           = - R^1{}_1 + \tilde R^{\dot 1}{}_{\dot 1} .\ee
The preserved symmetries at the boundary in this case thus form a copy of
\be sl(2|1) \times\widetilde{sl}(2|1)\ee
generated by
\beaa
& \mf L^{\alpha}{}_\beta\,, \qquad \mf R =
\alf\left( \mf D -J_{12} - J_{34} + J_{56}\right)\,,\nn\\
& \mf Q^\alpha = \mf Q^\alpha{}_3\,,\qquad       \mf S_{\alpha} = \mf S^{3}{}_\alpha\,,
\eeaa
and
\beaa
& \tilde{\mf L}^{\dot\alpha}{}_{\dot\beta}\,,\qquad
\tilde{\mf R} 
= \alf\left(\mf D -J_{12} -J_{34}-J_{56}\right)\,,
\nn\\
& \tilde{\mf Q}^{\dot\alpha}=\tilde{\mf Q}^{\dot \alpha 4}\,,
\qquad \tilde{\mf S}_{\dot\alpha} =\tilde{\mf S}_{4\dot\alpha}\,.
\eeaa

\subsection{Boundary degrees of freedom}
{\bf Boundary fields.}
The $\mathcal N=2$ fundamental hypermultiplet has as its field content a doublet of complex scalars $\phi^{\dot a}$ and two Weyl fermions $\psi_+^{\dot\alpha}$, $\psi_-^{\alpha}$. They transform as follows:
\be \begin{array}{l|ccccc}
       & sl(2) \times \widetilde{sl}(2) & J_{56} & so(1,3)      & \mf D  \\\hline
\phi   &   [0   ,\alf]          &  \phantom{+}0     &  [0 ,0]    &  1      \\
\psi_+ &   [0   ,   0]          & +\alf  &  [0 ,\alf]    & \textstyle{\frac{3}{2}} \\
\psi_- &   [0   ,   0]          & -\alf  &  [\alf, 0] & \textstyle{\frac{3}{2}}
\end{array}\label{d7fields}\ee

The fundamental matter fields listed in (\ref{d7fields}) are the basis of states of the rightmost site of the underlying spin chain. For each choice of vacuum, they fall into representations of the residual symmetry algebra labelled by the eigenvalues of $\mf D-J$. Of particular importance are the states with the lowest value of $\mf D - J$, which correspond in the scattering theory to possible orientations of the unexcited boundary.

~

\noindent
{\bf Bulk vacuum state $Z$.} In this case there is a degeneracy of states having the lowest possible value of $\mf D- J_{56}$, namely:
\be
\mf D - J_{56}=1\quad : \quad\phi^{\dot a}, \quad \psi_+^{\dot\alpha}\,.
\ee
These states transform in a fundamental representation $\boxslash=\bf(2|2)$ of $\widetilde{psl}(2|2)$, which therefore, from the point of view of the scattering theory, constitutes a degree of freedom carried by the boundary. The remaining orientations $\psi_-^{\alpha}$ have $\mf D- J_{56}=2$. They will participate in (the microscopic spin-chain description of) magnon scattering off the boundary, and possibly also in multiplets of boundary bound states.

Similarly, at the left-most site of the spin chain the conjugates of the fields in (\ref{d7fields}) appear, and the $\boxslash$ of states with $\mf D- J_{56}=1$ is spanned by $\bar\phi_{\dot a}$ and $\bar\psi_{-\dot\alpha}$.

This case is similar to that of the $Z=0$ giant gravitons in \cite{HMopen}, in the sense that the chain
carries boundary degrees of freedom. However, the left part of residual symmetry (\ref{D7zsim}) as well as the nature
of the boundary excitation is different. Consequently, the left factor of the boundary scattering matrix will be different.

~

\noindent
{\bf Bulk vacuum state $X$.}
Of the fields in the fundamental (\ref{d7fields}), there is a unique one, $\phi^{\dot 1}$, for which $\mf D- J_{12}$ is smallest (with $\mf D- J_{12} = \alf$). Similarly $\bar\phi_{\dot 2}$ is the lowest-lying anti-fundamental field. So in this case there are no degrees of freedom attached to the boundaries, and there is a unique unexcited configuration of the spin chain, namely
\be \bar \phi_{\dot 2} XXX\dots XXX\phi^{\dot 1}\, .\ee
As far as the scattering theory is concerned, this case is thus identical to that of the $Y=0$ giant gravitons in \cite{HMopen}, and the boundary reflection matrix will therefore be the same.

\subsection{Reflection matrices}
We can now determine the scattering matrix of a bulk magnon off the boundaries, for each of the choices of vacuum above. The bulk magnon transforms in a $(\boxslash, \boxslash)$ representation of the bulk symmetry $psl(2|2) \times \widetilde{psl}(2|2)\times \mathbb R^3$. Let us first collect the necessary facts about this representation.

Recall from \cite{Bsu22,Beis2006} that the representation $(\bf 2|2) = \boxslash$ is labelled by the values of the coefficients $a,b,c,d$ determining the action of the supersymmetries on the states,
\beaa & \mf Q^\alpha{}_a \ket{\phi^b} = a \delta_a^b \ket{\psi^\alpha}, \qquad\;\; &
    \mf Q^\alpha{}_a \ket{\psi^\beta} = b \eps^{\alpha\beta} \eps_{ab} \ket{\phi^b}\,, \\
&\mf S^a{}_\alpha \ket{\phi^b} = c \eps_{\alpha\beta} \eps^{ab} \ket{\psi^\beta}, \quad &
    \mf S^a{}_\alpha \ket{\psi^\beta} = d \delta_{\alpha}^\beta \ket{\phi^a}, \eeaa
and that these in turn depend on the momentum $p$ of the magnon according to
\be a=\sqrt{g} \eta\,,\quad b=\frac{\sqrt{g}}{\eta}f\left(1-\frac{x^+}{x^-}\right)\,,\quad
c=\frac{\sqrt{g}i \eta}{f x^+}\,,\quad d=\frac{\sqrt{g}}{i \eta}(x^+-x^-)\,, \label{qn} \ee
where $|\eta|^2=i(x^--x^+)$, to ensure unitarity, and $x^\pm$ are the standard spectral parameters
\be e^{i p} =  \frac{x^+}{x^-}\,, \qquad
x^+ +\frac{1}{x^+}-x^- -\frac{1}{x^-}=\frac{i}{g}\,. \label{sp}\ee
The second of these equations ensures that $a,b,c,d$ obey the condition $ad-bc=1$ for a short representation. The phase $f$ is the product $\prod_k e^{ip_k}$ over all magnons to the left of the magnon in question.

Equivalently, the representation is labelled by the values of the three central charges $C,P,K$ which occur in the brackets of the supersymmetries:
\beaa &&\left\{ \mf Q^\alpha{}_a , \mf Q^\beta{}_b\right\} = \eps^{\alpha\beta} \eps_{ab} P\,, \qquad
    \left\{ \mf S^a{}_\alpha , \mf S^b{}_\beta\right\} = \eps_{\alpha\beta} \eps^{ab} K\,,\nn
    \\
    &&\left\{ \mf Q^\alpha{}_a , \mf S^b{}_\beta\right\} = \delta_a^b \mf L^\alpha{}_\beta  \label{QSd7}
                                 +\delta^\alpha_\beta R^b{}_a + \delta^\alpha_\beta \delta_a^b C\,.\eeaa
They obey the shortening condition
\be C^2 - PK = \tfrac{1}{4}\, ,\ee
and are given in terms of the momenta by
\beaa
&& P = ab = gf\left(1- e^{ip} \right)  \,,\qquad K = cd = \frac{g}{f}\left(1-e^{-ip}\right)\,, \label{PK}
\\
&& C = \alf(ad+bc)= \alf\sqrt{1+16g^2 \sin(\tfrac{p}{2})^2}\,.\label{C}\eeaa

~

\noindent{\bf Vacuum $Z$}
As we found above, in this case the boundary transforms as
\be (1, \boxslash) \ee
with respect to the surviving symmetry
\be sl(2) \times sl(2) \times \widetilde{psl}(2|2)\times\mathbb R^3 .\ee
The reflection matrix therefore factors as a tensor product
\be \Refl \otimes \tilde{\Refl}\label{fullR}\ee
just as does the bulk S-matrix. Consider the untilded factor $\Refl$ first. In this factor the boundary scattering problem involves an excitation in a fundamental $\boxslash$ of $psl(2|2)\ltimes \mathbb R^3$ hitting the singlet boundary state and being reflected back into the bulk:
\be \Refl : \boxslash\otimes 1 \rightarrow \boxslash\otimes 1. \ee
The demand that $\Refl$ commute with the surviving $sl(2) \times sl(2)$ symmetry forces it to act as follows:
\beaa \Refl \ket{\phi^a_{p}} &=& M(p) \ket{\phi^a_{-p}} \nn\\
     \Refl \ket{\psi^\alpha_{p}} &=& N(p) \ket{\psi^\alpha_{-p}} \label{R1}\eeaa
for some functions $M(p)$, $N(p)$ of the incoming momentum $p$.
Here, of course, the absence of the supersymmetries $\mf Q^\alpha{}_a$ and $\mf S^a{}_\alpha$ means that the representation decomposes into the sum of two irreducible components,  $\boxslash \rightarrow \bf 2 \oplus \bf 2$, and symmetry arguments alone cannot fix the relative coefficient.

It is worth noting that the fact that the odd generators $\mf Q^\alpha{}_a$ and $\mf S^a{}_\alpha$ of $psl(2|2)$ are \emph{not} symmetries of the boundary is actually crucial. If they were then (\ref{QSd7}) would force the central charges $P$ and $K$ to depend on $p$ according (\ref{PK}). But then consider the scattering of a magnon off the right boundary. (The argument for the left boundary is similar.) The phase $f$ in (\ref{PK}) does not change, because it depends only on the other magnons, generically all far away to the left. Conservation of $P$ and $K$ would
then not allow $p\mapsto -p$ but only $p\mapsto p$, leaving us no sensible notion of reflection.
Note that the total values of all three central charges $C,P,K$ \emph{are} indeed conserved by reflections: they must be, because they occur in the brackets of (\ref{QSd7}) of the preserved supersymmetries $\tilde{\mf Q}^\alpha{}_a$, $\tilde{\mf S}^a{}_\alpha$. The point is simply that (\ref{PK}) is not valid for the untilded factor. And nor is (\ref{C}), which means that strictly speaking we have not yet shown that the outgoing momentum has to be $-p$; but this follows from the symmetries in the tilded factor $\tilde\Refl$ to be discussed below.

Given that symmetry alone does not completely determine $\Refl$, the natural question is whether there exist functions $M(p)$, $N(p)$ such that $\Refl(p)$ in (\ref{R1}) solves the boundary Yang-Baxter Equation (bYBE, and also known as the Reflection Equation)
\be \Smat(p,q)\,\Refl(p) \,\Smat(q,-p)\, \Refl(q)  = \Refl(q)\, \Smat(p,-q)\, \Refl(p)\, \Smat(-q,-p) \label{bYBE}\ee
which is the criterion for integrability in this context.
The bulk S-matrix acts in the following manner:
\beaa \Smat \ket{\phi_p^a \phi_q^b} &=& A(p,q) \ket{\phi_q^{\{a} \phi_p^{b\}}}
                                       +B(p,q) \ket{\phi_q^{ [a} \phi_p^{b ]}}
     + \alf C(p,q) \eps^{ab} \eps_{\alpha\beta}\ket{\psi_q^\alpha\psi_p^\beta} \nn\\
      \Smat \ket{\psi_p^\alpha \psi_q^\beta} &=& D(p,q) \ket{\psi_q^{\{\alpha} \psi_p^{\beta\}}}
                                       +E(p,q) \ket{\psi_q^{ [\alpha} \psi_p^{\beta ]}}
     + \alf F(p,q) \eps_{ab} \eps^{\alpha\beta}\ket{\phi_q^a\phi_p^b} \nn\\
      \Smat \ket{\phi_p^a \psi_q^\beta} &=& G(p,q) \ket{\psi_q^\beta \phi_p^a}
                                       +  H(p,q) \ket{\phi_q^a \psi_p^\beta } \nn\\
      \Smat \ket{\psi_p^\alpha \phi_q^b} &=& K(p,q) \ket{\psi_q^\alpha \phi_p^b}
                                       +  L(p,q) \ket{\phi_q^b \psi_p^\alpha }\,,\eeaa
where \cite{Beis2006}
{\small \beaa A(p_1,p_2) &=& S_0(p_1,p_2) \frac{x_2^+- x_1^-}{x_2^- - x_1^+} \nn\\
      B(p_1,p_2) &=& S_0(p_1,p_2) \frac{x_2^+ - x_1^-}{x_2^- - x_1^+}
            \left( 1 - 2 \frac{1 - 1/x_2^- x_1^+}{1 - 1/x_2^+ x_1^+}\,\,
                         \frac{ x_2^- - x_1^-}{x_2^+ - x_1^-}\right) \nn\\
C(f, p_1,p_2) &=&  \frac{2}{f} S_0(p_1,p_2) \frac{ \eta_1 \eta_2}{ x_1^+ x_2^+} \frac{ x_2^- - x_1^-}{x_2^- - x_1^+} \frac{1}{1 - 1/x_2^+ x_1^+} \nn\\
D(p_1,p_2) &=& - S_0(p_1,p_2)\nn\\
E(p_1,p_2) &=& - S_0(p_1,p_2)\left( 1 -  2 \frac{ 1 - 1/x_2^+ x_1^-}{1 - 1/x_2^- x_1^-}\,\,
                             \frac{ x_2^+ - x_1^+}{x_2^- - x_1^+}\right)\nn\\
F(f,p_1,p_2) &=& - 2f S_0(p_1,p_2) \frac{(x_1^+ - x_1^-) (x_2^+ - x_2^-)}{\eta_1 \eta_2 x_1^- x_2^-}\,
         \frac{x_2^+ - x_1^+}{x_2^- - x_1^+}\,\frac{1}{1 - 1/x_2^- x_1^-} \nn\\
G(p_1,p_2) &=&  S_0(p_1,p_2)\frac{x_2^+ - x_1^+}{x_2^- - x_1^+}\nn\\
H(p_1,p_2) &=&  S_0(p_1,p_2)\frac{\eta_1}{\eta_2} \frac{ x_2^+ - x_2^-}{ x_2^- - x_1^+} \nn\\
K(p_1,p_2) &=&  S_0(p_1,p_2)\frac{\eta_2}{\eta_1} \frac{ x_1^+ - x_1^-}{ x_2^- - x_1^+} \nn\\
L(p_1,p_2) &=& S_0(p_1,p_2)\frac{x_2^- - x_1^-}{x_2^- - x_1^+}\,. \label{Smatfns}\eeaa}

Certain components of (\ref{bYBE}) are entirely diagonal and hold solely by virtue of the invariance of the relevant functions in $\Smat$ under the parity transformation
\be (p_1, p_2) \leftrightarrow (-p_2,-p_1) \qquad \Longleftrightarrow \qquad x_1^\pm  \leftrightarrow -x_2^\mp\,.\ee
For example (\ref{bYBE}) is correct acting on $\ket{\phi_p^1 \phi_q^1}$ provided that $A(p,q) = A(-q,-p)$, which is indeed true (assuming the overall factor $S_0(p,q)$ is also parity-invariant).
Then there are certain matrix elements of (\ref{bYBE}) which contain non-diagonal contributions but which again hold purely by parity invariance of the S-matrix. For example
\be \bra{\phi^1_{-p}\psi^+_{-q}} \text{(bYBE)} \ket{\phi^1_p\psi^+_q} \ee
holds by virtue of
\be H(p,q)=K(-q,-p) \quad\text{and}\quad G(p, q) = L(-q, -p) .\ee
In the end it turns out that (\ref{bYBE}) encodes essentially only two independent constraints on $M(p), N(p)$. First, the matrix element
\be \bra{\psi^+_{-p}\phi^1_{-q}} \text{(bYBE)}\ket{\phi^1_p\psi^+_q} \ee
yields the equation
\beaa 0&=& N(p) N(q) G(-q,-p) H(p,-q)  - N(p) M(q) G(q,-p) H(p,q) \nn\\ &&{} + M(p) N(q) G(p,-q) K(-q,-p) - M(p) M(q) G(p,q) K(q,-p)\, .\eeaa
On substituting for $G,H,K$ from (\ref{Smatfns}) one finds that it is possible to separate variables and solve this equation by setting
\be \frac{N(p) x^-(p) + M(p) x^+(p)}{N(p) - M(p)}  = x_B \label{sep}\ee
where for the moment $x_B$ can be any constant. Thus
\be M/N = \frac{x_B-x^-}{x_B + x^+}\,. \ee
But there is a second constraint, which occurs in the matrix element
\be \bra{\phi^1_{-p}\phi^2_{-q}} \text{(bYBE)}\ket{\psi^+_p\psi^-_q} \ee
and which is solved only by using the mass-shell condition (\ref{sp}) \emph{and} in addition imposing the equation
\be x_B + \frac{1}{x_B} = \frac{i}{g} \,.\label{xBeqn}\ee
We have verified that, given this equation, all the remaining components of the boundary Yang-Baxter equation are satisfied. We have therefore the most general form of reflection matrix consistent with integrability:
\beaa \Refl \ket{\phi^a_{p}} &=& R_0(p) \left(\frac{x_B-x^-}{x_B + x^+} \right)\ket{\phi^a_{-p}} \nn\\
      \Refl \ket{\psi^\alpha_{p}} &=& R_0(p) \ket{\psi^\alpha_{-p}} \label{R2}\eeaa
for some function $R_0(p)$.

Since we used the bYBE in arriving at this result, we cannot strictly deduce that this is the correct reflection matrix: integrability is merely a consistent assumption rather than an outcome in this case.
Nevertheless, the forms of the reflection matrix (\ref{R2}) and relation (\ref{xBeqn}) are very natural in light of what happens in the other factor $\tilde\Refl$ of the full reflection matrix (\ref{fullR}). Here the boundary scattering problem involves the bulk excitation in a fundamental $\boxslash$ of $\widetilde{psl}(2|2)\ltimes \mathbb R^3$ reflecting off a boundary degree of freedom in another fundamental $\boxslash$:
\be
\tilde\Refl : \boxslash \otimes \boxslash \rightarrow \boxslash \otimes \boxslash \,.
\ee
Since we found that this process must respect the full $\widetilde{psl}(2|2)\ltimes \mathbb R^3$ symmetry, the situation in this factor is identical to that of the $Z=0$ case in \cite{HMopen}. The reader is referred to that paper for the full details. In particular, one expects that the boundary degree of freedom transforms in the representation defined by the following coefficients:
\be a_B = \sqrt g \eta_B\,, \quad b_B = \frac{\sqrt g f_B}{\eta_B} \,,\quad
 c_B = \frac{\sqrt g i \eta_B}{x_B f_B}\,, \quad d_B= \frac{\sqrt g x_B}{i \eta_B} \label{bndryabcd}\ee
where $\left|\eta_B\right|^2=-ix_B$, $f_B$ is the boundary phase, and the mass-shell condition $ad-bc=1$ yields precisely the relation (\ref{xBeqn}) that was needed for the bYBE to hold in the untilded factor. The positive energy solution is
\be x_B= \frac{i}{2g} \left( 1 + \sqrt{1 + 4 g^2}\right) \ee
and implies that the energy of the unexcited boundary is
\be \epsilon = ad+bc = \sqrt{1+4g^2}\, . \ee
The anomalous dimension of an operator with no bulk excitations should therefore be
\be
2\left(\sqrt{1+4g^2}-1\right) = 4 g^2+{\cal O}(g^4)
= \frac{\lambda}{4\pi^2} +{\cal O}(\lambda^2)\,.\ee
The quantity $\tfrac{\lambda}{4\pi^2}=\tfrac{g_{YM}^2 N}{4 \pi^2}$ is
precisely the one-loop anomalous dimension of an operator of the
form $\bar\phi ZZ\dots ZZ \phi$, as can easily be read from
the computations of \cite{ErMa}. (When the vacuum is $Z=X_5 + iX_6$, the quantities $R_1$ and $R_L$ of equation
(8) in \cite{ErMa} vanish.)

The boundary scattering matrix for a right boundary in the tilded factor is
\beaa \tilde\Refl \ket{\phi_p^a \phi_B^b} &=& A_R(p,q) \ket{\phi_{-p}^{\{a} \phi_B^{b\}}}
                                       +B_R(p,q) \ket{\phi_{-p}^{ [a} \phi_B^{b ]}}
     + \alf C_R(p,q) \eps^{ab} \eps_{\alpha\beta}\ket{\psi_{-p}^\alpha\psi_B^\beta} \nn\\
      \tilde\Refl \ket{\psi_p^\alpha \psi_B^\beta} &=& D_R(p,q) \ket{\psi_{-p}^{\{\alpha} \psi_B^{\beta\}}}
                                       +E_R(p,q) \ket{\psi_{-p}^{ [\alpha} \psi_B^{\beta ]}}
     + \alf F_R(p,q) \eps_{ab} \eps^{\alpha\beta}\ket{\phi_{-p}^a\phi_B^b} \nn\\
      \tilde\Refl \ket{\phi_p^a \psi_B^\beta} &=& G_R(p,q) \ket{\psi_{-p}^\beta \phi_B^a}
                                       +  H_R(p,q) \ket{\phi_{-p}^a \psi_B^\beta } \nn\\
      \tilde\Refl \ket{\psi_p^\alpha \phi_B^b} &=& K_R(p,q) \ket{\psi_{-p}^\alpha \phi_B^b}
                                       +  L_R(p,q) \ket{\phi_{-p}^b \psi_B^\alpha }\,, \eeaa
where
\beaa A_R(p) &=& \tilde R_0(p) \frac{x^- \left(x^--x_B\right)}{x^+ \left(x^++x_B\right)}\,,\nn\\
      B_R(p) &=&  \tilde R_0(p)\frac{x^- \left(-2 \left(x^-\right)^2+x^+
   x^-+2 \left(x^+\right)^2\right)-x_B \left(2 \left(x^-\right)^2+x^+ x^--2
   \left(x^+\right)^2\right)}{\left(x^+\right)^2 \left(x_B+x^+\right)}\,,\nn\\
      C_R(p) &=& -\tilde R_0(p) \frac{2\eta_B\eta(x^+,x^-)}{f} \frac{
   \left(x_B+x^--x^+\right) \left(x^-+x^+\right) }{ x^+
   \left(x_B+x^+\right)}\,,\nn\\
      D_R(p) &=&\tilde R_0(p)\,, \nn\\
      E_R(p) &=&\tilde R_0(p)\frac{x^+ \left(2 \left(x^-\right)^2+x^+ x^--2 \left(x^+\right)^2\right)+x_B
   \left(-2 \left(x^-\right)^2+x^+ x^-+2 \left(x^+\right)^2\right)}{x^- x^+ \left(x_B+x^+\right)}\,,\nn\\
      F_R(p) &=&\tilde R_0(p) \frac{2f}{\eta_B \eta(x^+,x^-)} \frac{
\left(\left(x^-\right)^2-\left(x^+\right)^2\right) \left(x^- x^++x_B \left(x^+-x^-\right)\right)}{x^- \left(x^+\right)^2 \left(x_B+x^+\right)}\,,\nn\\
      G_R(p)&=&  -\tilde R_0(p)\frac{\eta(x^+,x^-)}{\eta_B} \frac{x_B
   \left(x^-+x^+\right)}{ x^+
   \left(x_B+x^+\right)}\,,\nn\\
      H_R(p) &=&\tilde R_0(p)\frac{\left(x^+\right)^2-x_B x^-}{x^+
   \left(x_B+x^+\right)}\,,\nn\\
     K_R(p) &=&\tilde R_0(p)\frac{\left(x^-\right)^2+x_B x^+}{\left(x^+\right)^2+x_B
   x^+}\,,\nn\\
     L_R(p) &=&\tilde R_0(p) \frac{\eta_B}{\eta(x^+,x^-)} \frac{\left(x^--x^+\right) \left(x^-+x^+\right)}{x^+ \left(x_B+x^+\right) }\,,\eeaa
which satisfies the bYBE, and of course will coincide with the right-boundary reflection matrix of case $Z=0$ in \cite{HMopen}.

~

\noindent{\bf Vacuum $X$} As we showed above, for the other choice of vacuum, $X=X_1+iX_2=\phi^{14}$ the symmetries and degrees of freedom are similar to those of the strings ending on $Y=0$ giant gravitons in \cite{HMopen}. Let us briefly
review the derivation of the boundary scattering matrix and show that is consistent with the 1-loop results obtained in \cite{ErMa}, according to which the left reflection amplitudes for each of the scalar field impurities are,
\be
R_Y = e^{ip}\,,\qquad R_{\bar Y} = e^{-ip}\,, \qquad R_{Z} = R_{\bar Z} = -1\,.
\label{1loopD7}
\ee

To characterize the reflection of bulk magnons it is convenient to understand the preserved $sl(2|1) \times\widetilde{sl}(2|1)$
as a subalgebra of the bulk symmetry algebra corresponding to the vacuum $X$. In terms of $sl(2)\times \widetilde{sl}(2)\cong so(4)_{3456}$ spinorial indices, the preserved supersymmetry generators (\ref{D7Xsusys})
are\footnote{The generators $R^{a}_{~b}$ and $\tilde{R}^{\dot a}_{~\dot b}$ of $so(4)_{3456}$ can be taken
\be \begin{array}{ll}
 R^1{}_2 = \mf R^2{}_3 & \tilde R^{\dot 1}{}_{\dot 2} = -\mf R^4{}_1 \\
 R^2{}_1 = \mf R^3{}_2 & \tilde R^{\dot 2}{}_{\dot 1} = -\mf R^1{}_4 \\
 R^1{}_1 = - R^2{}_2 = \alf \mf R^2{}_2 - \alf \mf R^3{}_3\qquad
& \tilde R^{\dot 1}{}_{\dot 1} = - \tilde R^{\dot 2}{}_{\dot 2} = \alf \mf R^4{}_4 - \alf \mf R^1{}_1\end{array} \nn\ee}
\be
\mf Q^\alpha{}_{2}\,, \quad \tilde{\mf Q}^{\dot \alpha}{}_{\dot 2}\,,
\quad \mf S^{2}{}_\alpha\,, \quad \tilde{\mf S}^{\dot 2}{}_{\dot\alpha}\,.
\label{D7Xsusysb}
\ee
Once again, the reflection matrix factors into the form
\be
\Refl \otimes \tilde{\Refl}\,.\label{fullR1}
\ee
Commutation with the bosonic generators requires the reflection matrix to be diagonal. For a left reflection, commutation with fermionic
generators (\ref{D7Xsusysb}) fixes each of the factors in (\ref{fullR1}) to be of the form
\be
{\cal R}_L=\tilde{\cal R}_L = \left(\begin{array}{cccc}
r_1 & 0& 0& 0\\
0 & r_2 &0 &0\\
0 & 0 & r & 0\\
0 & 0 &  0&r
\end{array}\right) =
R_0(p)\left(\begin{array}{cccc}
- e^{ip} & 0& 0& 0\\
0 & 1 &0 &0\\
0 & 0 & 1 & 0\\
0 & 0 &  0& 1
\end{array}\right)
\label{RLD7X}
\ee
To compare with the 1-loop results (\ref{1loopD7}), we need to write
the scalar fields carrying $sl(2)\times \widetilde{sl}(2)\cong so(4)_{3456}$ spinorial indices,
\be
Y=\phi^{1}\times\tilde\phi^{\dot 1}\,,\quad \bar Y=\phi^{2}\times\tilde\phi^{\dot 2}\,,\quad
Z=\phi^{1}\times\tilde\phi^{\dot 2}\,,\quad \bar Z=\phi^{2}\times\tilde\phi^{\dot 1}\,.
\ee
We observe that all the relative exact reflection amplitudes from (\ref{RLD7X})
are consistent with the 1-loop results (\ref{1loopD7}).

Again, the boundary scattering matrix (\ref{RLD7X}) coincides with that of the
case $Y=0$ of \cite{HMopen} and bYBE is therefore obeyed.

\section{D5-brane}\label{sec:D5}
\label{D5symme}
\subsection{Boundary symmetries}
Consider now a D5-brane whose worldvolume wraps an $AdS_4\subset AdS_5$ and a maximal $S^2\subset S^5$. For this case, instead of
fixing the orientation of the brane and considering different choices for the R-charge of the vacuum state, we will fix the bulk vacuum state to be $Z$
and consider different orientations for the maximal $S^2$.
The original $so(6)$ R-symmetry is broken by the presence of the D5-brane down to $so(3)_H \times so(3)_V$. We will consider the following two inequivalent situations:\footnote{Taking e.g.  $X^1=X^2=X^6=0$ is less interesting because, with the vacuum state $Z$, both $so(3)_H$ and $so(3)_V$ would be broken.}

({\it i})  ~  Maximal $S^2$ specified by $X^4=X^5=X^6=0$.

({\it ii}) ~  Maximal $S^2$ specified by $X^1=X^2=X^3=0$.

\noindent
The bosonic symmetries preserved by these two brane configurations are identical.
Of the Lorentz generators, only $M_{01}$, $M_{02}$ and $M_{12}$ will be preserved, which form a diagonal $so(1,2)$:
\begin{align}
&L^{\check +}_{~\check -} = \mathfrak{L}^{+}_{~-} + \tilde{\mathfrak{L}}^{\dot-}_{~\dot+}\,,\nn
\\
& L^{\check -}_{~\check +} = \mathfrak{L}^{-}_{~+} + \tilde{\mathfrak{L}}^{\dot+}_{~\dot-}\,,\nn
\\
&  L^{\check +}_{~\check +} = \mathfrak{L}^{+}_{~+} + \tilde{\mathfrak{L}}^{\dot-}_{~\dot-}\,,
\end{align}
where the $L^{\check\alpha}_{~\check\beta}$ follow canonical commutation rules.
Of the original $so(6)$ generators only $J_{12}$, $J_{13}$, $J_{23}$,
$J_{45}$, $J_{46}$ and $J_{56}$ will be preserved. These can be written
as two sets of canonical $sl(2)$ generators $R^{a}_{~b}$ and
$\tilde{R}^{\dot a}_{~\dot b}$:
\begin{eqnarray}
&R^{{1}}_{~ {2}} =  {\mathfrak{R}}^{1}_{~2} - {\mathfrak{R}}^{4}_{~3}\,,
\qquad  \qquad   \qquad   \qquad
&\tilde{R}^{\dot1}_{~\dot2} =  {\mathfrak{R}}^{1}_{~4} - {\mathfrak{R}}^{2}_{~3}\,,\nn
\\
& R^{2}_{~1} =  {\mathfrak{R}}^{2}_{~1} - {\mathfrak{R}}^{3}_{~4}\,,
\qquad \qquad  \qquad \qquad
& \tilde{R}^{\dot2}_{~\dot1} =  {\mathfrak{R}}^{4}_{~1} - {\mathfrak{R}}^{3}_{~2}\,,\nn
\\
&  R^{1}_{~1} = \tfrac12( {\mathfrak{R}}^{1}_{~1} - {\mathfrak{R}}^{2}_{~2}
- {\mathfrak{R}}^{3}_{~3} + {\mathfrak{R}}^{4}_{~4})\,,
\quad
& \tilde{R}^{\dot1}_{~\dot1} = \tfrac12( {\mathfrak{R}}^{1}_{~1} + {\mathfrak{R}}^{2}_{~2}
- {\mathfrak{R}}^{3}_{~3} - {\mathfrak{R}}^{4}_{~4})\,,
\end{eqnarray}
In the case ({\it i}), the $R^{a}_{~b}$ and the $\tilde{R}^{\dot a}_{~\dot b}$ give rise to
$so(3)_H$ and $so(3)_V$ respectively. These roles are exchanged in the case ({\it ii}).

Both D-brane configurations, ({\it i}) and ({\it ii}), preserve half the background supersymmetries. The preserved combinations, which can be written as carrying indices of the preserved
$so(1,2)$, $so(3)_H$ and $so(3)_V$, turn out to be (see appendix \ref{D5susy})
\be
\begin{array}{ll}
 Q^{\check \pm}_{~1\dot1} = \mathfrak{Q}^\pm_{~1} - \kappa \tilde{\mathfrak{Q}}^{\dot\mp3} \,,
\qquad\qquad  &
 S_{~~\check \pm}^{1\dot1} = \mathfrak{S}^1_{~\pm} - \tfrac1\kappa \tilde{\mathfrak{S}}_{3\dot\mp} \,,
\\
 Q^{\check \pm}_{~1\dot2} = \mathfrak{Q}^\pm_{~4} - \kappa \tilde{\mathfrak{Q}}^{\dot\mp2} \,,
\qquad\qquad &
 S_{~~\check \pm}^{1\dot2} = \mathfrak{S}^4_{~\pm} - \tfrac1\kappa \tilde{\mathfrak{S}}_{2\dot\mp} \,,
\\
 Q^{\check \pm}_{~2\dot1} = \mathfrak{Q}^\pm_{~2} - \kappa \tilde{\mathfrak{Q}}^{\dot\mp4} \,,
\qquad\qquad &
 S_{~~\check \pm}^{2\dot1} = \mathfrak{S}^2_{~\pm} - \tfrac1\kappa \tilde{\mathfrak{S}}_{4\dot\mp}\,,
\\
 Q^{\check \pm}_{~2\dot2} = -\mathfrak{Q}^\pm_{~3} + \kappa \tilde{\mathfrak{Q}}^{\dot\mp1} \,,
\qquad\qquad &
S_{~~\check \pm}^{2\dot2} = -\mathfrak{S}^3_{~\pm} + \tfrac1\kappa \tilde{\mathfrak{S}}_{1\dot\mp} \,,
\end{array}
\label{QSosp}
\ee
where $\kappa=i$ in case ({\it i}) and $\kappa=1$ in case ({\it ii}).

~

\noindent {\bf Bulk vacuum state $Z$.} The choice of vacuum state
\be Z= X_5 + i X_6 \ee
breaks the $so(3)_{456}$ symmetry generated by $\tilde{R}^{\dot a}_{~\dot b}$, which is $so(3)_V$ in the case ({\it i}) and $so(3)_H$ in the case ({\it ii}).
Among the supersymmetries (\ref{QSosp}) of the D5-brane,  $Q^{\check\alpha}_{~a\dot 2}$ and $S^{a\dot 2}_{~~\check\alpha}$ are charged under $\mf D-J_{56}$ and do not preserve $Z$.
This leaves
\be L^{\check\alpha}_{~\check\beta},\quad R^{a}_{~b},\quad
Q^{\check\alpha}_{~a\dot 1} ,\quad S^{a\dot 1}_{~~\check\alpha}\ee
as residual symmetries of both the boundary and the vacuum.
Since
\be \{Q_{~a\dot 1}^{\check\alpha},{S}^{b\dot 1}_{~~\check\beta} \}=
\delta_{a}^{b} {L}_{~\check\beta}^{\check\alpha} +\delta_{\check\beta}^{\check\alpha} R_{~a}^{b} +
\delta_{\check\beta}^{\check\alpha}\delta_{a}^{b} (\mf D - J_{56}) ,\ee
the boundary symmetries certainly include a diagonal $sl(2|2)_D$ subalgebra of the bulk symmetry algebra $psl(2|2)\otimes\widetilde{psl}(2|2)\ltimes \mathbb R^3$, with
\be C_D = \mathfrak{D}-J_{56}  \,.\label{CD} \ee
Interestingly, the presence or absence of additional non-vanishing central charges, $P_D$ and $K_D$, depends on the choice of brane orientation. One has
\be \{Q_{~a\dot 1}^{\check\alpha},Q_{~b\dot 1}^{\check\beta} \}=
\eps^{\check\alpha\check\beta} \eps_{ab} (P-\kappa^2 \tilde{P})\,,
\qquad
\{S^{a\dot 1}_{~~\check\alpha},S^{b\dot 1}_{~~\check\beta} \}=
\eps_{\check\alpha\check\beta} \eps^{ab} (K-\tfrac{1}{\kappa^2} \tilde{K})\,,\label{QQSS}\ee
and since untilded and tilded central charges are identified, $P=\tilde P, K=\tilde K$, the additional central charges are twice the bulk additional central charges in the case ({\it i}), whereas they vanish in the case ({\it ii}). This will be important for the consistency of reflection processes in what follows.

\subsection{Boundary degrees of freedom}

\noindent
{\bf Boundary fields.}
The 3d hypermultiplet living on the defect has as its field content an $so(3)_H$-doublet of complex bosonic scalars and an $so(3)_V$-doublet of 3d fermionic spinors:
\be \begin{array}{l|cccc}
       & so(3)_H \times so(3)_V & so(1,2)      & \mf D  \\\hline
\phi^a   &      [\alf   ,   0]       &    [0]    &  \alf      \\
\psi^{\check \alpha\dot a} &     [0   ,\alf]        &   [\alf]    & 1
\end{array}\label{d5fields}\ee

~

\noindent
{\bf Case ({\it i}): $so(3)_V$ broken.} We now ask which of the fundamental matter fields, which will occupy the right-most site of the underlying spin chain, have the lowest possible value of $\mf D- J_{56}$. In this case $\phi^a$ are not charged under  $J_{56}$, while
the $\psi^{\check\alpha\dot a}$ have charges $\pm\tfrac12$. So the lowest-lying fields are
\be
\mf D - J_{56}=\tfrac12 \quad : \quad\phi^{ a}, \quad \psi^{\check\alpha\dot1}\,.
\ee
These transform in a fundamental representation $\boxslash=\bf(2|2)$ of ${psl}(2|2)_D$.  We need to determine the parameters $(a,b,c,d)$ specifying this representation. We expect that they should correspond to radial line segments in the LLM disk picture \cite{HM,LLM}, and they should certainly yield an expression for the energy that matches the known 1-loop results of \cite{DeMa}. To achieve this we take\footnote{ Note that this parameterization and the resulting definitions of $x_B$ and $\eta_B$ differ from the D7 case of the previous section.}
\begin{equation}
a = \sqrt{2g} \eta_B\,,
\quad
b =  \frac { \sqrt{2g}f_B}{\eta_B}\,,
\quad
c = \frac{\sqrt{2g} i \eta_B}{x_B f_B}\,,
\quad
d =  \frac{ \sqrt{2g} x_B}{i \eta_B}\,,
\label{abcdBD5}
\end{equation}
where $f_B$ is a phase giving the starting point of a radial line segment on the rim of the
unit disk (for a right boundary).
The unitarity and shortening conditions give
\begin{equation}
|\eta_B|^2 = -i x_B\,, \qquad  x_B \equiv \frac{i(1+\sqrt{1+16 g^2})}{4g} \, .
\end{equation}
The central charge associated with the energy of the boundary excitation is
\begin{equation}
C_D = \mathfrak{D}-J_{56}  =\frac{1}{2} \sqrt{1+16 g^2}\,.
\label{exacte}
\end{equation}
Then in the weak coupling limit $\epsilon \approx  \tfrac12 + 4 g^2$. We  can consider
a bosonic boundary excitation in order to compare with the 1-loop anomalous
dimension calculations of  DeWolfe and Mann \cite{DeMa}. The $\tfrac12$ represents
the classical dimension of the boundary scalar field, while the $4 g^2$ matches exactly
half of the 1-loop anomalous dimension of an operator of the form $\bar \phi Z\dots Z \phi $
(where our $Z$ is made out of $X_V$ according to the conventions of \cite{DeMa}).

~

\noindent {\bf Case ({\it ii}): $so(3)_H$ broken.}
Now, fields $\phi^{\dot a}$ have charges $\pm\tfrac12$ under $J_{56}$, while
the $\psi^{\check\alpha a}$ are uncharged . Thus, the lowest possible
value of $\mf D- J_{56}$ is
\be
\mf D - J_{56}=0 \quad : \quad\phi^{\dot 1}\,.
\ee
Setting this field in the right-most site of the spin chain, the right boundary would carry no degree of freedom,
{\it i.e.} the right-most site is occupied by a singlet of ${psl}(2|2)_D$.

~

\subsection{Bulk degrees of freedom and reflection matrices}
As we have seen, only a diagonal $psl(2|2)_D\times \mathbb R^3$ of the bulk symmetry $psl(2|2) \times \widetilde{psl}(2|2)\times \mathbb R^3$ is preserved by the boundaries. We distinguished two cases, depending on the relative orientation of the vacuum and the D5-brane in the internal space. We must now determine how bulk magnons are accommodated into representations of the preserved ${psl}(2|2)_D\times \mathbb R^3$.

With respect to the bulk symmetry, bulk excitations transform in a product of a fundamental $(\phi^a|\psi^\alpha)$ of $psl(2|2)\ltimes \mathbb R^3$ and a fundamental $(\tilde\phi^{\dot a}|\tilde\psi^{\dot \alpha})$ of $\widetilde{psl}(2|2)\ltimes \mathbb R^3$.
By acting with the diagonal generators, one can see that $(\phi^a|\psi^\alpha)$
transforms also in the fundamental of  $psl(2|2)_D\times \mathbb R^3$, with labels $(a,b,c,d)$
given by (\ref{qn}). Analogously, $(\tilde\phi^{\dot a}|\tilde\psi^{\dot \alpha})$ also transforms in the fundamental of  $psl(2|2)_D\times \mathbb R^3$, when reorganized as $(\tilde\phi^{\dot 1},\tilde\phi^{\dot 2}|\tilde\psi^{\dot -},\tilde\psi^{\dot +})$ and with labels  $(-\kappa a, \kappa b,\tfrac{c}{\kappa},-\tfrac{d}{\kappa})$.

Therefore the bulk magnons transform in the following tensor product of fundamental representations
of the diagonal symmetry (following the notation of \cite{Beis2006}):
\be
\langle 0,0;C,P,K\rangle \otimes \langle0,0;C,-\kappa^2 P,-\tfrac{1}{\kappa^2}K\rangle
= \{0,0;2C,(1-\kappa^2 )P,(1-\tfrac{1}{\kappa^2})K\} \,,
\ee

~

\noindent
{\bf Case ({\it i}): $so(3)_V$ broken.} Taking $\kappa=i$ we get for the diagonal central charges
\be
C_D=2C\,,\qquad P_D =2P\,,\qquad K_D=2K\,,
\ee
which satisfy the multiplet splitting condition \cite{Beis2006,CDO}
\begin{equation}
{C_D}^2 - {P_D} K_D = 1\,,
\end{equation}
according to which
\be
\{0,0;2C,2P,2K\} = \langle1,0;2C,2P,2K\rangle \oplus \langle1,0;2C,2P,2K\rangle = \boxslash\hspace{-.18cm}\boxslash \oplus \twobv \,.
\ee

As we saw above, the right boundary carries a $\boxslash$ spanned by the fields  $\phi^{ a}$ and $ \psi^{\check\alpha\dot1}$.
Therefore we shall be interested in the following two scattering processes:
\ba
&& \Refl : \boxslash\hspace{-.18cm}\boxslash \otimes\;\boxslash \rightarrow \boxslash\hspace{-.18cm}\boxslash \otimes \boxslash
\label{R8x4}
\\
&& \Refl : \twobv \otimes \boxslash \rightarrow \twobv \otimes \boxslash
\label{R8x4p}
\ea

Similar processes were studied in \cite{AF2008}, for the bulk scattering of elementary magnon against bound states of magnons.
In general, tensor products of short representations can have more than one irreducible component. For example, for $m,n>1$,
\be
\langle m,n;\vec C\rangle \otimes \langle 0,0;\vec{C}'\rangle  = \{m,n;\vec C+\vec{C}'\} \oplus \{m-1,n-1;\vec C+\vec{C}'\}\,.
\ee
However, the tensor products in the scattering processes (\ref{R8x4}) and (\ref{R8x4p}) still have a single irreducible
component
\ba
&& \langle 1,0;\vec C\rangle \otimes \langle0,0;\vec{C}'\rangle  = \{1,0;\vec C+\vec{C}'\} \,,
\label{irre8x4}
\\
&& \langle 0,1;\vec C\rangle \otimes \langle0,0;\vec{C}'\rangle  = \{0,1;\vec C+\vec{C}'\}  \,.
\label{irre8x4p}
\ea
Thus, by demanding that ${\cal R}$ commute with the  generators of the residual symmetry, we will be
able to fix each of the boundary scattering matrices (\ref{R8x4}) and (\ref{R8x4p}) up to an overall factor.
Let us first focus on the reflection by a right boundary (\ref{R8x4}). To commute with the bosonic
generators, the ${\cal R}$ matrix  has to be of the form
\ba
{\cal R}|\phi_p^{bc},\phi_B^a\rangle &\!\!= \!\!&
A_{1}(p) |\phi_{-p}^{\{ bc},\phi_{B}^{ a\}}\rangle + A_2(p) \epsilon_{de}
\epsilon^{a\{b}|\phi_{-p}^{c\}e},\phi_B^d\rangle +A_{11}(p) \epsilon_{\calpha\cbeta}
|\phi_{-p}^{\calpha\cbeta},\phi_B^{\{b}\rangle \epsilon^{c\}a}\nn
\\
&&
+A_{13}(p) \epsilon_{\calpha\cbeta}
\epsilon^{a\{b}|\psi_{-p}^{c\}\cbeta},\psi_B^\calpha\rangle
\nn\\
{\cal R}|\phi_p^{\calpha\cbeta},\phi_B^a\rangle &\!\!= \!\!&
A_{6}(p) |\phi_{-p}^{\calpha\cbeta},\phi_B^{a}\rangle
+ A_{15}(p)|\psi_{-p}^{a[\calpha},\psi_B^{\cbeta]}\rangle
+A_{10}(p)\epsilon_{bc}\epsilon^{\calpha\cbeta}|\phi_{-p}^{ca},\phi_B^b\rangle
\nn\\
{\cal R}|\psi_p^{b\cbeta},\phi_B^a \rangle &\!\!= \!\! &
A_{3}(p) |\psi_{-p}^{\{b\cbeta},\phi_B^{a\}}\rangle+
A_{4}(p) |\psi_{-p}^{[b\cbeta},\phi_B^{a]}\rangle
+A_{16}(p) \epsilon^{ab}\epsilon_{\cgamma\cdelta}
      |\phi_{-p}^{\cgamma\cdelta},\psi_B^{\cbeta}\rangle\nn\\
&&+A_{19}(p)|\phi_{-p}^{ab},\psi_B^\cbeta\rangle
\nn\\
{\cal R}|\phi_p^{ab},\psi_B^\calpha \rangle &\!\!= \!\! &
A_{5}(p) |\phi_{-p}^{ab},\psi_B^\calpha \rangle+
A_{18}(p) |\psi_{-p}^{\{b\calpha},\phi_B^{a\}}\rangle
\nn\\
{\cal R}|\phi_p^{\cbeta\cgamma},\psi_B^\calpha\rangle &\!\!= \!\!&
A_{9}(p) |\phi_{-p}^{\cbeta\cgamma},\psi_B^{\calpha}\rangle
+A_{17}(p)\epsilon_{ab}\epsilon^{\cbeta\cgamma}|\psi_{-p}^{b\calpha},\phi_B^a\rangle
\nn\\
{\cal R}|\psi_p^{b\cbeta},\psi_B^\calpha \rangle &\!\!= \!\! &
A_{7}(p) |\psi_{-p}^{b\{\cbeta },\psi_B^{\calpha\}}\rangle +
A_{8}(p) |\psi_{-p}^{b[\cbeta },\psi_B^{\calpha]}\rangle
+A_{12}(p) \epsilon^{\calpha\cbeta} \epsilon_{cd}|\phi_{-p}^{bd},\phi_B^c\rangle\nn\\
&&+ A_{14}(p) \epsilon^{\calpha\cbeta} \epsilon_{\cgamma\cdelta}|\phi_{-p}^{\cgamma\cdelta},\phi_B^{b}\rangle
\label{As}.
\ea

\begin{figure}
\begin{center}
\epsfxsize=4.3in\leavevmode\epsfbox{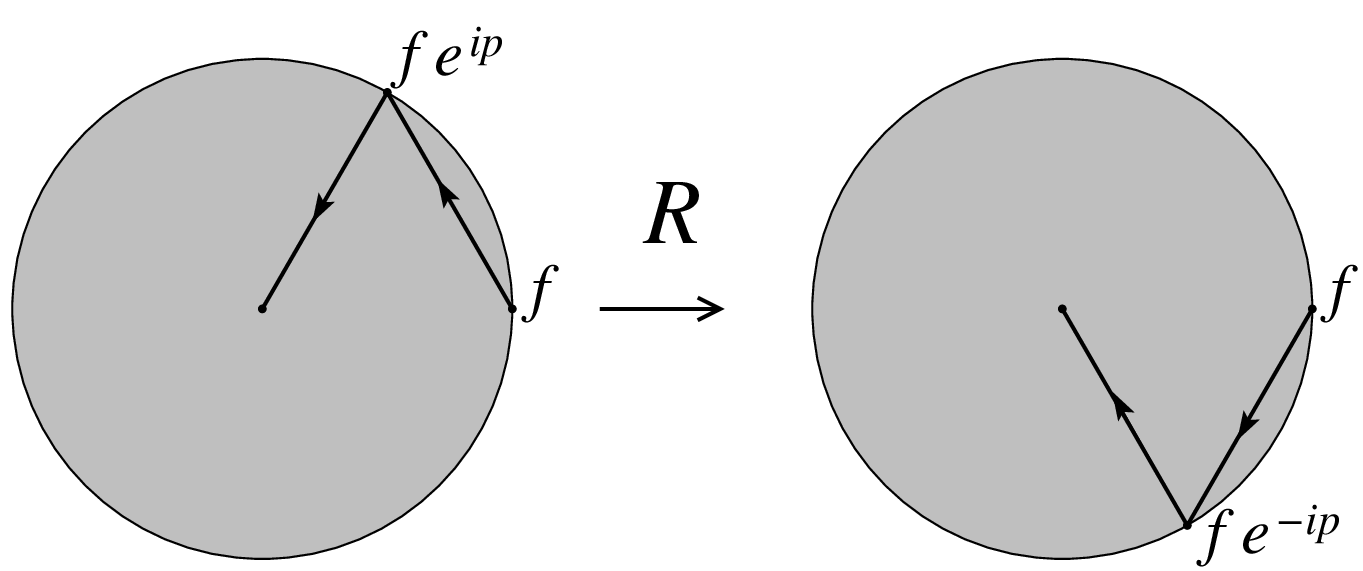}
\end{center}
\caption{Scattering by a right boundary with a fundamental degree of freedom.} \label{fig1}
\end{figure}

The vanishing of the commutator with the fermionic generators fixes 18 of these arbitrary functions, leaving unknown only an overall factor. We list the results explicitly in appendix \ref{AppABs}. To compute these commutators, one must know the quantum labels of excitations before and after the reflection. Let us call $f$ the starting point of the bulk magnon in the cumulative picture. So, the $\twobh$ representation labels are
\be
a=\sqrt{g} \eta\,,\quad
b=\frac{\sqrt{g}}{\eta}f\left(1-\frac{x^+}{x^-}\right)\,,\quad
c=\frac{\sqrt{g}i \eta}{f x^+}\,,\quad
d=\frac{\sqrt{g}}{i \eta}(x^+-x^-)\,.
\ee
Then, the starting point of the boundary excitation is $f_B=f e^{ip}=f\tfrac{x^+}{x^-}$.
So, its labels are
\begin{equation}
a_B = \sqrt{2g} \eta_B\,,
\quad
b_B =  \frac {\sqrt{2g}f}{\eta_B}\frac{x^+}{x^-}\,,\quad c_B = \frac{\sqrt{2g} i \eta_B}{x_B
f} \frac{x^-}{x^+}\,,\quad d_B =  \frac{ \sqrt{2g} x_B}{i \eta_B}\,.
\end{equation}
After the scattering, the bulk excitation has reversed its momentum, so the line representing it in the cumulative picture has also to be  reversed. When doing so, the net central charges have to be conserved. Then, the cumulative picture has to change as shown in figure \ref{fig1}, which means that the representation labels change in the following way
\be
a'=a \,,\quad
b'= -\tfrac{x^-}{x^+} b \,,\quad
c'= -\tfrac{x^+}{x^-} c\,,\quad
d'=d\,.
\ee
\begin{equation}
a_B' = a_B \,,\quad
b_B' = \left(\tfrac{x^-}{x^+}\right)^2 b_B\,,\quad
c_B' =  \left(\tfrac{x^+}{x^-}\right)^2 c_B \,,\quad
d_B' =  d_B\,,
\end{equation}

Analogously, for the the ${\cal R}$ matrix of (\ref{R8x4p}) we obtain
\ba
{\cal R}|\phi_p^{\cbeta\cgamma}, \psi_B^\calpha\rangle &\!\!= \!\!&
B_{1}(p) |\phi_{-p}^{\{ \cbeta\cgamma},\psi_B^{ \calpha\}}\rangle + B_2(p) \epsilon_{\cdelta\varepsilon}
\epsilon^{\calpha\{\cbeta}|\phi_{-p}^{\cgamma\}\varepsilon},\psi_B^\cdelta\rangle
+B_{11}(p) \epsilon_{ab} |\phi_{-p}^{ab},\psi_B^{\{\cgamma}\rangle \epsilon^{\cbeta\}\calpha}\nn
\\
&&
+ B_{13}(p) \epsilon_{ab} \epsilon^{\calpha\{\cbeta}|\psi_{-p}^{b\cgamma\}},\phi_B^a\rangle
\nn\\
{\cal R}|\phi_p^{ab},\psi_B^\calpha\rangle &\!\!= \!\!&
B_{6}(p) |\phi_{-p}^{ab},\psi_B^{\calpha}\rangle
+ B_{15}(p) |\psi_{-p}^{[a\calpha},\phi_B^{b]}\rangle
+B_{10}(p)\epsilon_{\cbeta\cgamma}\epsilon^{ab}|\phi_{-p}^{\cgamma\calpha},\psi_B^\cbeta\rangle
\nn\\
{\cal R}|\psi_p^{b\cbeta},\psi_B^\calpha \rangle &\!\!= \!\! &
B_{3}(p) |\psi_{-p}^{b\{\cbeta },\psi_B^{ \calpha\}}\rangle+
B_{4}(p) |\psi_{-p}^{b[\cbeta },\psi_B^{ \calpha]}\rangle
+B_{16}(p) \epsilon^{\calpha\cbeta}\epsilon_{cd} |\phi_{-p}^{cd},\phi_B^{b}\rangle
\nn\\
&& +B_{19}(p)|\phi_{-p}^{\calpha\cbeta},\phi_B^b\rangle
\nn\\
{\cal R}|\phi_p^{\calpha\cbeta},\phi_B^a \rangle &\!\!= \!\! &
B_{5}(p) |\phi_{-p}^{\calpha\cbeta} ,\phi_B^a\rangle+
B_{18}(p) |\psi_{-p}^{a \{\cbeta},\psi_B^{\calpha\}}\rangle
\nn\\
{\cal R}|\phi_p^{bc},\phi_B^a\rangle &\!\!= \!\!&
B_{9}(p) |\phi_{-p}^{bc},\phi_B^{a}\rangle
+B_{17}(p)\epsilon_{\calpha\cbeta}\epsilon^{bc}|\psi_{-p}^{a\cbeta},\psi_B^\calpha\rangle
\nn\\
{\cal R}|\psi_p^{b\cbeta},\phi_B^a\rangle &\!\!= \!\! &
B_{7}(p) |\psi_{-p}^{\{b \cbeta },\phi_B^{a\}}\rangle+
B_{8}(p) |\psi_{-p}^{[b \cbeta}, \phi_B^{a]}\rangle
+B_{12}(p) \epsilon^{ab} \epsilon_{\cgamma\cdelta}|\phi_{-p}^{\cbeta\cdelta},\psi_B^\cgamma\rangle
\nn\\
&&+B_{14}(p) \epsilon^{ab} \epsilon_{cd}|\phi_{-p}^{cd},\psi_B^{\cbeta}\rangle
\ea
for coefficient functions $B_i(p)$ given in appendix \ref{AppABs}.

Symmetry arguments thus determine the form of the reflection matrix for a single bulk magnon off the right boundary up to two unknown functions, which specify the overall factors for (\ref{R8x4}) and (\ref{R8x4p}). The important question is then whether there exists any choice of these functions, or more precisely their ratio, such that the system is integrable, i.e. such that the complete reflection matrix obeys the boundary Yang-Baxter equation.

We have found that in fact there is no choice of relative factor between the processes (\ref{R8x4}) and (\ref{R8x4p}) such that all components of the bYBE are satisfied\footnote{See the note added at the end of the Introduction.}. This failure is related to the fact that the bulk S-matrix (\ref{Smatfns}) does not respect the decomposition of the bulk magnons
\be \boxslash \otimes \boxslash \rightarrow \boxslash\hspace{-.18cm}\boxslash \oplus \twobv\ee
into their graded-symmetric and graded-antisymmetric parts: the `C' and `F' channels mix these in non-trivial ways.
As we note below, this puts into context the known results about integrability at 1-loop \cite{DeMa} and the failure to prove integrability for classical open strings on D5-branes \cite{Mann}.

~

\noindent {\bf Case ({\it ii}): $so(3)_H$ broken.}
We consider now the case $\kappa=1$ in (\ref{QQSS}). The diagonal central charges for the bulk
excitations are then
\be
C_D = \sqrt{1+16g^2 \sin(\tfrac{p}{2})^2}\,,
\qquad P_D=0\,, \qquad K_D = 0\,,
\label{CPKii}
\ee
which, for generic $p$, do not obey any shortening condition. Then, the 16 bulk magnons
transform in the smallest long representation $\{0,0;\vec C_D\}$ of $psl(2|2)_D\times \mathbb R^3$.
At the right boundary is the field  $\phi^{\dot 1}$,  which is a singlet of the
preserved $psl(2|2)_D\times \mathbb R^3$. Since neither the boundary nor the bulk excitation carry the additional central charges $P_D$ and $K_D$, the conservation of these charges imposes no constraints on the reflection matrix.

To obtain the values of $C_D,P_D,K_D$ above, bulk excitations transform as a tensor product of two fundamental excitations of $psl(2|2)_D\times \mathbb R^3$, with labels $(a,b,c,d)$ and  $(-a,b,c,-d)$ respectively. The reflection matrix is therefore a map
\be
\Refl : \boxslash\otimes\boxslash\otimes 1 \rightarrow \boxslash\otimes\boxslash\otimes 1
\ee
and is fixed by the bosonic symmetries to be of the form
\ba
\Refl \;|\phi_p^a \times \tilde\phi_p^b \rangle \!&=& \!
A_R(p) |\phi_{-p}^{\{a} \times \tilde \phi_{-p}^{b\}} \rangle +B_R(p) |\phi_{-p}^{ [a} \times \tilde\phi_{-p}^{b ]}\rangle
+ \alf C_R(p) \eps^{ab} \eps_{\calpha\cbeta} |\psi_{-p}^\calpha \times\tilde \psi_{-p}^\cbeta \rangle \nn
\\
\Refl \;|\psi_p^\calpha \times \tilde\psi_p^\cbeta \rangle \!&=& \!
D_R (p)|\psi_{-p}^{\{\calpha}\times \tilde\psi_{-p}^{\cbeta\}}\rangle
+ E_R(p) |\psi_{-p}^{ [\calpha} \times \tilde\psi_{-p}^{\cbeta ]}\rangle
+ \alf F_R(p) \eps_{ab} \eps^{\calpha\cbeta}|\phi_{-p}^a \times\tilde\phi_{-p}^b\rangle
\nn\\
\Refl \;|\phi_p^a \times \tilde\psi_p^\cbeta\rangle \!&=& \!
G_R(p) |\psi_{-p}^\cbeta\times\tilde \phi_{-p}^a\rangle + H_R(p) |\phi_{-p}^a \times\tilde \psi_{-p}^\cbeta\rangle
\nn\\
\Refl \;|\psi_p^\calpha \times\tilde \phi_p^b\rangle \!&=& \!
K_R(p) |\psi_{-p}^\calpha \times \tilde\phi_{-p}^b\rangle +  L_R(p) |\phi_{-p}^b \times\tilde \psi_{-p}^\calpha\rangle \,.
\ea
After the scattering the quantum labels change to  $(a',b',c',d')$ and  $(-a',b',c',-d')$
where
\be
a'=a \,,\quad
b'= -\tfrac{x^-}{x^+} b \,,\quad
c'= -\tfrac{x^+}{x^-} c\,,\quad
d'=d\,.
\ee
The commutation of $\Refl$ with the fermionic generators requires
\beaa
A_R(p) \!&=& \!   - R_0(p) \frac{x^-}{x^+}\,,
\nn\\
B_R(p) \!&=& \!   R_0(p) \frac{x^-(x^-+(x^+)^3)}{(x^+)^2(1+x^-x^+)}\,,
\nn\\
C_R(p) \!&=& \!  -R_0(p) \frac{\eta^2(x^-+x^+)}{f x^+(1+x^-x^+)}\,,
\nn\\
D_R(p) \!&=& \!  R_0(p) \,,
\nn\\
E_R(p)\!&=& \!  - R_0(p)  \frac{x^++(x^-)^3}{x^-(1+x^-x^+)} \,,
\nn\\
F_R(p)\!&=& \!   R_0(p) \frac{f(x^-+x^+)(x^- -x^+)^2}{\eta^2 x^+(1+x^-x^+)} \,,
\nn\\
G_R(p)\!&=& \!    R_0(p)  \frac{x^- + x^+}{2x^+}\,,
\nn\\
H_R(p)\!&=& \!   R_0(p) \frac{x^+ -x^-}{2x^+}\,,
\nn\\
K_R(p)\!&=& \!    R_0(p) \frac{x^+ -x^-}{2x^+}\,,
\nn\\
L_R(p)\!&=& \!   R_0(p)\frac{x^- +x^+}{2x^+}\,.
\label{D5iicoef}
\eeaa
We would like to compare these reflection amplitudes with those computed at 1-loop
for scalar fields in \cite{DeMa}. Of course, since the all-loop expressions  are known up to an overall factor, we should
compare relative amplitudes between different scalar fields. According to the 1-loop calculation,
when the scalar fields  $\Phi_I$ with $I=1,2,3$ and $\Phi_4$ are reflected by a right boundary,
they pick up the following factors \cite{DeMa}
\beaa
\Refl \;|\Phi_I(p)\rangle \!&=& \! - e^{-ip}|\Phi_I(-p)\rangle\nn
\\
\Refl \;|\Phi_4(p)\rangle \!&=& \! |\Phi_4(-p)\rangle
.\eeaa
In our notation, the scalars $\Phi_I$ correspond to $\phi^{\{a} \times \tilde \phi^{b\}}$
while $\Phi_4$ corresponds to $\phi^{[a} \times \tilde \phi^{b]}$ and their all-loop reflection
amplitudes are given by $A_R(p)$ and $B_R(p)$ respectively. Expanding in powers of $g$, one sees that indeed
\be
\frac{A_R(p)}{B_R(p)}  =  - \frac{x^+ +x^-(x^+)^2}{x^-+(x^+)^3}  =   - e^{-ip} +{\cal O}(g^3)\,,
\ee
where we have used that
\be x^{\pm} =  e^{\pm i \tfrac p2}\ \frac{1+\sqrt{1+16g^2
\sin^2(\tfrac p2)}}{4 g \sin(\tfrac p2)}\,, \qquad\quad \lambda =
16\pi^2 g^2\,. \ee

Now, as with case ({\it i}) above,  the reflection
amplitudes (\ref{D5iicoef}) turn out not to satisfy the boundary
Yang-Baxter equation\footnote{See the note added at the end of the Introduction.}. A direct computation shows that many matrix
elements of the bYBE are non-vanishing. For instance,
\beaa && \hspace{-1.5cm}\langle (\phi^{1} \times \tilde
\phi^{2})_{-p_1} ,(\phi^{1} \times \tilde \phi^{1})_{-p_2} |
\text{(bYBE)} |(\phi^{1} \times \tilde \phi^{2})_{p_1} ,(\phi^{1}
\times \tilde \phi^{1})_{p_2}\rangle =\nn\\
&& \hspace{-.8cm}
\frac{x_2^-(x_1^+-x_1^-)^2(x_1^++x_1^-)(x_2^+-x_2^-)(x_2^++x_2^-)(x_1^+-x_2^-)(x_1^--x_2^-)(x_1^-+x_2^+)}{4
x_1^+ (x_2^+)^2(x_1^--x_2^+)^2(x_1^+ + x_2^+)^2(1+x_1^+
x_1^-)(1-x_1^+ x_2^+)}\;, \label{nvbybe} \eeaa
The 1-loop analysis carried out in \cite{DeMa} shows that in the
scalar sector of theory, the fields living in the defect yield
integrable open boundary conditions for the 1-loop dilatation
operator. Therefore, consistency with this result requires that any
non-vanishing matrix element of the bYBE for scalar bulk excitations
should vanish in the weak coupling limit. We have verified that this is indeed the case. For example the weak coupling expansion of (\ref{nvbybe}) gives
\beaa && \hspace{-2.5cm}\langle (\phi^{1} \times \tilde
\phi^{2})_{-p_1} ,(\phi^{1} \times \tilde \phi^{1})_{-p_2} |
\text{(bYBE)} |(\phi^{1} \times \tilde \phi^{2})_{p_1} ,(\phi^{1}
\times \tilde \phi^{1})_{p_2}\rangle = \nn\\
&& \hspace{-2.3cm}  - \frac{256 g^2e^{-\tfrac i2 (3p_1+p_2)}(1-2
e^{i p_1}+e^{i(p_1+p_2)})\cos(\tfrac {p_1}2)\sin^4(\tfrac
{p_1}2)\cos(\tfrac {p_2}2)\sin^2(\tfrac
{p_2}2)\sin(\tfrac{p_1+p_2}{2})\sin(\tfrac{p_1-p_2}{2})}{(1+e^{2i
p_2}(5-4\cos(p_1))+2e^{i p_2}(\cos(p_1)-2))^2}\nn\\
\hspace{-.3cm}+{\cal O}(g^3)\,,
 \label{nvbybeg} \eeaa
From this analysis it is clear that the one-loop integrability in of the D5 brane system is ``an accident'' (and one which will not be exclusive to the scalar sector, since all the components of bYBE are order $g^2$). The breakdown of integrability beyond one-loop is consistent with the lack of integrability in the classical boundary conditions on the string side \cite{Mann}.

\section{Conclusions}\label{conc}
Let us summarize our results concerning the reflection matrices for the scattering of elementary magnons from boundaries in the open spin chains associated to the D7 and D5 gauge theories. In each case we found that symmetry arguments alone are sufficient to determine the reflection matrices up to at most two unknown functions.

For the D7 theory, the reflection matrix is (depending on the choice of vacuum, $X$ or $Z$) either integrable automatically, with only one overall factor undetermined, or integrable given a certain apparently natural choice of the ratio between the two unknowns.  The remaining overall factor can naturally be constrained by demanding crossing symmetry, in a way entirely analogous to \cite{ChCo,ABR}. The survival of integrability seems to be closely linked to the fact that the boundary respects the factorization $psl(2|2) \times \widetilde{psl}(2|2)$ of bulk scattering processes (which was also true in the giant graviton case \cite{HMopen}).

In contrast, in the D5 theory we have shown that boundary scattering is certainly not integrable. There exists no reflection matrix satisfying the boundary Yang Baxter-Equation consistent with the symmetries of the problem\footnote{See the note added at the end of the Introduction.}. This breakdown of integrability is not visible at leading order at weak coupling, essentially because it is linked to the bulk scattering processes associated to length-changing interactions of the spin chain.

It would be interesting to investigate all-loop reflection matrices for other situations in which open spin chains have arisen in an AdS/CFT context. One notable example is that of supersymmetric Wilson loops with operator insertions. In \cite{DK} the study of 1-loop anomalous dimension
of certain scalar insertions is reduced to an open spin chain with integrable boundary conditions.
The Wilson loops preserve a copy of $osp(2,2|4)$  \cite{GreenEtAl} which is the same superalgebra (though differently embedded in $psl(4|4)$) preserved by the D5 branes we consider. The two situations can therefore be expected to show some similarities, but further work is needed to determine whether or not the boundary Yang-Baxter can be satisfied in the Wilson loop case.

 \bigskip

\textit{Acknowledgments.-- } We thank Peter Bowcock, Heng-Yu Chen, Patrick Dorey, Nick Dorey, Dan Freedman, David Kagan, Anshuman Maharana, Julian Sonner and Douglas Smith for helpful discussions. D.H.C. is funded by the Seventh Framework Programme under grant agreement number PIEF-GA-2008-220702. C.A.S.Y. is funded by the Leverhulme trust.

\appendix
\section{Conventions}\label{conventions}
The vector representation $\bf 6$ of $so(6)$ is equivalent to the antisymmetric second-rank tensor representation of $sl(4) \cong so(6)$; to translate between them we make a standard choice
\begin{align} &X = \Phi_1 + i\Phi_2 = \phi^{14}\,,&  \bar X = \Phi_1 - i \Phi_2 = \phi^{23}\,, \nn\\
 &Y = \Phi_3 + i\Phi_4 = \phi^{24}\,,& \bar Y = \Phi_3 - i \Phi_4 = \phi^{31}\,, \nn\\
 &Z = \Phi_5 + i\Phi_6 = \phi^{34}\,,&  \bar Z = \Phi_5 - i \Phi_6 = \phi^{12}\, .\end{align}
This corresponds to the following set of $so(6)$ gamma matrices:
\begin{align} &\Gamma_1 = -\sigma_2 \otimes \sigma_2 \otimes \sigma_3\,, &\Gamma_2 = \sigma_2 \otimes \sigma_1 \otimes 1\,,\nn\\
& \Gamma_3 = -\sigma_2 \otimes \sigma_2 \otimes \sigma_1\,, &\Gamma_4 = \sigma_2 \otimes \sigma_2 \otimes \sigma_2\,, \nn\\
&\Gamma_5 = \sigma_1 \otimes 1\otimes 1\,, & \Gamma_6 = -\sigma_2 \otimes \sigma_3 \otimes 1\,, \end{align}
for then
\be \Gamma_7 = i \Gamma_1 \dots \Gamma_6 = \sigma_3 \otimes 1 \otimes 1,\qquad  C = \sigma_2 \otimes 1 \otimes \sigma_2  \ee
and one may verify that when ${\bf 6 \rightarrow 4 \otimes 4}$ according to $\Phi_i \mapsto \half (1+\Gamma_7) \Gamma_i C$, the identifications above are obtained.

For the  $so(1,3)$ gamma matrices we use:
\begin{align}
&\gamma^0 = -i\sigma_1 \otimes 1, &\gamma^i = \sigma_2 \otimes \sigma_i\,,
\end{align}
Alternatively the same basis can be written as:
\be
\gamma^\mu = \left(\begin{array}{cc}
0 & -i\sigma^\mu
\\
i\bar\sigma^\mu & 0
\end{array}\right)\quad {\rm with}\quad
\begin{array}{l}
\sigma^\mu = (1,{\vec{\sigma}})
\\
\bar\sigma^\mu = (-1,{\vec{\sigma}})
\end{array}
\ee
\be \gamma_5 = i\gamma_0\gamma_1\gamma_2\gamma_2 = \sigma_3 \otimes 1 =
\left(\begin{array}{cc}
\delta^\alpha_\beta & 0
\\
0 & -\delta^{\dot\beta}_{\dot\alpha}
\end{array}\right),\quad  C =-i \sigma_3 \otimes \sigma_2 =
\left(\begin{array}{cc}
\epsilon_{\alpha\beta} & 0
\\
0 & \epsilon^{\dot\alpha\dot\beta}
\end{array}\right),   \ee

For the  $so(1,9)$ gamma matrices one can use:
\begin{align}
& \Gamma^\mu = \gamma^\mu \otimes 1, & {\rm for\ } \mu=0,\ldots 3 \,,
\label{g03}
\\
& \Gamma^I = \gamma_5 \otimes \Gamma^{I-3}, & {\rm for\ } I=4,\ldots 9 \,.
\label{g49}
\end{align}

\section{D5-brane supersymmetries}
\label{D5susy}

The supersymmetries preserved by the D5-branes considered in section \ref{D5symme}
can be worked out by looking for the Killing spinors of the supersymmetric background
consistent with the kappa symmetry projection \cite{ST}. From the field theory point of view,
the original supersymmetry transformations need to be restricted to those preserving
the position of the defect. Ignoring the $R$-charge indices of of the supersymmetry generators,
supertranslations $\eps \cdot Q$ preserving the position $x_3=0$ of a defect
would be those satisfying $\eps= \eps\ \gamma^1\gamma^2\gamma^3$.
However, the original CFT is ${\cal N}=4$ and the generators carry $su(4)$ indices.
The $R$-charge indices on the preserved combinations will depend on the specific $S^2$ that
the D5-brane wraps. These combinations can be elucidated repeating the analysis of \cite{ST}, for the two D5-branes
configurations we are interested in. For the supersymmetries at least, one can also consider the D3-D5 brane intersections in $d=10$ Minkowski space. In order to satisfy the kappa symmetry projection, the Killing spinors have to be projected as
\be
P_+ \varepsilon = \varepsilon\qquad {\rm with}\qquad
P_{+} = \left\{
\begin{array}{ll}
 \tfrac12(1+ \Gamma^{3456})  & {\rm for\ the\ case\ } (i)\,,
\\
 \tfrac12(1+ \Gamma^{3789})  & {\rm for\ the\ case\ } (ii)\,,
\end{array}    \right.
 \ee
where $ \Gamma^A$ are the $SO(1,9)$ Dirac matrices. Using conventions (\ref{g03})-(\ref{g49}) for them,
this  projector is  reduced to
\be
P_+ = \half \left(1+ \gamma^0\gamma^1\gamma^2 \otimes i\Gamma_1 \Gamma_2 \Gamma_3\right)
= \half \left(1 + \sigma_1\otimes\sigma_3 \otimes  \sigma_2 \otimes \sigma_1 \otimes \sigma_2\right)\,.
\label{123}
\ee
in the case ({\it i}) and  to
\be
P_+ = \half \left(1+ \gamma^0\gamma^1\gamma^2 \otimes i\Gamma_4 \Gamma_5 \Gamma_6\right)
=  \half \left(1- \sigma_1\otimes\sigma_3 \otimes  \sigma_1 \otimes \sigma_1 \otimes\sigma_2\right)\,.
\label{456}
\ee
in the case ({\it ii}).

Now, in order to match with the preserved supersymmetries in the dual field theory,
it is convenient to regard the supersymmetry generators of ${\cal N}=4$ SYM as the 32 components
of an object $Q$ in
\be
Dirac_{SO(1,3)} \times Dirac_{SO(6)}\,,
\ee
for which we should allow only supertranslations $\eps \cdot Q$ such that the Lorentz and $SO(6)$
chiralities match:
\be
\eps = \eps (\gamma_5 \otimes \Gamma_7) \label{eQ} \,.
\ee
The relations between the original ${\mf Q}^\alpha_{\ {\rm a}}$ and $\tilde {\mf Q}^{\dot\alpha {\rm a}}$
and the $\eps \cdot Q$ subject to (\ref{eQ}) are the following. For Lorentz indices
\ba
&& {\mf Q}_{- {\rm a}} = {\mf Q}^{+}_{\ {\rm a}}  =(\up,\up,\ldots)\cdot Q\,,
\qquad
-i\tilde{{\mf Q}}^{\dot- {\rm a}} = i\tilde{{\mf Q}}_{\dot+}^{\ {\rm a}} =(\down,\up,\ldots)\cdot Q\,,
\label{qLor1}\nn
\\
-\!\!\!\!\!\!\!\!\!\!&& {\mf Q}_{+ {\rm a}} =  {\mf Q}^{-}_{\ {\rm a}}  = (\up,\down,\ldots)\cdot Q\,,
\qquad\,\,\,\,\,
i\tilde{{\mf Q}}^{\dot+{\rm a}} = i\tilde{{\mf Q}}_{\dot-}^{\ {\rm a}}=(\down,\down,\ldots)\cdot Q\,.
\label{qLor2}
\ea
For $su(4)$ indices
\ba
&& {\mf Q}^\alpha_{\ 1} = (\ldots,\up,\up,\up)\cdot Q\,, \qquad    i \tilde{{\mf Q}}^{\dot\alpha 1} =  -(\ldots,\down,\up,\down)\cdot Q\,,
\label{qsu1}\nn
\\
&& {\mf Q}^\alpha_{\ 2} = (\ldots,\up,\up,\down)\cdot Q\,, \qquad    i\tilde{{\mf Q}}^{\dot\alpha 2} =
(\ldots,\down,\up,\up)\cdot Q\,,
\label{qsu2}\nn
\\
&& {\mf Q}^\alpha_{\ 3} = (\ldots,\up,\down,\up)\cdot Q\,, \qquad    i \tilde{{\mf Q}}^{\dot\alpha 3} = - (\ldots,\down,\down,\down)\cdot Q\,,\nn
\label{qsu3}
\\
&& {\mf Q}^\alpha_{\ 4} = (\ldots,\up,\down,\down)\cdot Q\,, \qquad
i\tilde{{\mf Q}}^{\dot\alpha 4} = (\ldots,\down,\down,\up)\cdot Q\,.
\label{qsu4}
\ea

We should treat the superconformal transformations $S$ accordingly, {\it i.e.}
as a 32 component object  provided we allow only transformations $\eta \cdot S$
such that
\be
\eta = - \eta (\gamma_5 \otimes \Gamma_7) \,.
\label{eS}
\ee
Again, superconformal generators ${\mf S}^{{\rm a}}_{\alpha}$ and $\tilde {\mf S}_{{\rm a}\dot\alpha}$
can be related to $\eta \cdot S$. For that, the same identifications with undotted, dotted, upstairs and
downstairs indices as in (\ref{qLor1})-(\ref{qsu4}) hold.
One can verify that with these identifications, the superbrackets taking the form
\ba
\left\{Q_{ir},Q_{js}\right\}  \!\! & = & \!\!  2 P_\mu (\gamma^\mu C)_{ij} \bar C_{rs}\,,\nn
\\
\left\{S_{ir},S_{js}\right\}  \!\! & = & \!\! - 2 K_\mu (\gamma^\mu C)_{ij} \bar C_{rs}\,,\nn
\\
\left\{Q_{ir} , S_{js}\right\}  \!\! & = & \!\!
C_{ij}(\Gamma^{ab}\bar C)_{rs} J_{ab}+(\gamma^{\mu\nu}C)_{ij}\bar C_{rs}M_{\mu\nu}+C_{ij}\bar C_{rs}D\,,
\label{QS}
\ea
where $J_{ab}$ are the generators of $so(6)$ and $M_{\mu\nu}$ of $so(1,3)$,
are translated to
\ba
\{\mathfrak{Q}^{\alpha}_{~\rm b},\tilde{\mathfrak{Q}}^{\dot\alpha{\rm a}}\}
\!\! & = & \!\! 2 P_\mu  (\sigma^\mu)^{\alpha\dot\alpha} \delta^{\rm a}_{\rm b}\,,\nn
\\
\{\tilde{\mathfrak{S}}_{\dot\alpha{\rm a}},\mathfrak{S}_{\alpha}^{~\rm b}\}
\!\! & = & \!\! 2 K_\mu  (\bar\sigma^\mu)_{\dot\alpha\alpha} \delta_{\rm a}^{\rm b}\,,\nn
\\
\{\mathfrak{S}^{\rm a}_{~\alpha},\mathfrak{Q}^{\beta}_{~\rm b}\}
\!\! & = & \!\! \delta^{\rm a}_{\rm b} \mathfrak{L}^\beta_{~\alpha}
+  \delta^\beta_{\alpha}  {\mathfrak{R}}^{\rm a}_{~\rm b}
+\tfrac12 \delta^{\rm a}_{\rm b} \delta^\beta_{\alpha} \mathfrak{D} \,,\nn
\\
\{\tilde{\mathfrak{S}}_{{\rm a}\dot\alpha},\tilde{\mathfrak{Q}}^{\dot\beta\rm b}\}
\!\! & = & \!\!\delta_{\rm a}^{\rm b}\tilde{\mathfrak{L}}^{\dot\beta}_{~\dot\alpha}
-  \delta^{\dot\beta}_{\dot\alpha}  {\mathfrak{R}}^{\rm b}_{~\rm a}
+\tfrac12 \delta^{\rm b}_{\rm a} \delta^{\dot\beta}_{\dot\alpha} \mathfrak{D}\, .
\ea

Therefore, among the $\eps \cdot Q$ satisfying (\ref{eQ}) and  $\eta \cdot S$ satisfying (\ref{eS}),
the supersymmetries of  the dCFT are found demanding also $\eps = \eps P_+$ and $\eta = \eta P_+$.
In the case ({\it i}), where $P_+$ is given by (\ref{123}), the preserved combinations are:
\ba
 &(\up,\upd,\up,\up,\up) \cdot Q \mp (\down,\upd,\down, \down, \down )\cdot Q =
 {\mf Q}^{\pm}_{\ 1} - i\tilde{{\mf Q}}^{\dot\mp 3}\nn
 \\
&(\up,\upd,\up,\up,\down) \cdot Q \pm(\down,\upd,\down, \down, \up )\cdot Q =
 {\mf Q}^\pm_{\ 2} - i\tilde{{\mf Q}}^{\dot\mp 4}\nn\\
 &(\up,\upd,\up,\down,\up) \cdot Q \mp (\down,\upd,\down, \up, \down )\cdot Q =
 {\mf Q}^\pm_{\ 3} - i\tilde{{\mf Q}}^{\dot\mp 1}\nn\\
&(\up,\upd,\up,\down,\down) \cdot Q \pm (\down,\upd,\down, \up, \up )\cdot Q =
 {\mf Q}^\pm_{\ 4} - i \tilde{{\mf Q}}^{\dot\mp 2}\nn\\
 &(\down,\upd,\up,\up,\up) \cdot S \mp (\up,\upd,\down, \down, \down )\cdot S =
 i \tilde{{\mf S}}_{1\dot\pm} + {{\mf S}}^{3}_{\ \mp}\nn
 \\
&(\down,\upd,\up,\up,\down) \cdot S \pm(\up,\upd,\down, \down, \up )\cdot S =
 i \tilde{{\mf S}}_{2\dot\pm} + {{\mf S}}^{4}_{\ \mp}\nn
 \\
 &(\down,\upd,\up,\down,\up) \cdot S \mp (\up,\upd,\down, \up, \down )\cdot S =
 i \tilde{{\mf S}}_{3\dot\pm} + {{\mf S}}^{1}_{\ \mp}\nn
 \\
&(\down,\upd,\up,\down,\down) \cdot S \pm (\up,\upd,\down, \up, \up )\cdot S =
 i \tilde{{\mf S}}_{4\dot\pm} + {{\mf S}}^{2}_{\ \mp}
\ea
Whereas in the case ({\it ii}), with $P_+$ given by (\ref{456}), we obtain:
\ba
 &(\up,\upd,\up,\up,\up) \cdot Q \pm i (\down,\upd,\down, \down, \down )\cdot Q =
 {\mf Q}^{\pm}_{\ 1} - \tilde{{\mf Q}}^{\dot\mp 3}\nn
 \\
&(\up,\upd,\up,\up,\down) \cdot Q \mp i (\down,\upd,\down, \down, \up )\cdot Q =
 {\mf Q}^\pm_{\ 2} - \tilde{{\mf Q}}^{\dot\mp 4}\nn\\
 &(\up,\upd,\up,\down,\up) \cdot Q \pm i (\down,\upd,\down, \up, \down )\cdot Q =
 {\mf Q}^\pm_{\ 3} - \tilde{{\mf Q}}^{\dot\mp 1}\nn\\
&(\up,\upd,\up,\down,\down) \cdot Q \mp i(\down,\upd,\down, \up, \up )\cdot Q =
 {\mf Q}^\pm_{\ 4} -  \tilde{{\mf Q}}^{\dot\mp 2}\nn\\
 &(\down,\upd,\up,\up,\up) \cdot S \pm i (\up,\upd,\down, \down, \down )\cdot S =
 i \tilde{{\mf S}}_{1\dot\pm} -i {{\mf S}}^{3}_{\ \mp}\nn
 \\
&(\down,\upd,\up,\up,\down) \cdot S \mp i (\up,\upd,\down, \down, \up )\cdot S =
 i \tilde{{\mf S}}_{2\dot\pm} -i {{\mf S}}^{4}_{\ \mp}\nn
 \\
 &(\down,\upd,\up,\down,\up) \cdot S \pm i (\up,\upd,\down, \up, \down )\cdot S =
 i \tilde{{\mf S}}_{3\dot\pm} -i {{\mf S}}^{1}_{\ \mp}\nn
 \\
&(\down,\upd,\up,\down,\down) \cdot S \mp i (\up,\upd,\down, \up, \up )\cdot S =
 i \tilde{{\mf S}}_{4\dot\pm} -i {{\mf S}}^{2}_{\ \mp}
\ea
These are the combinations presented in (\ref{QSosp}), with $\kappa=i$ for the case ({\it i})
and $\kappa=1$ for the case ({\it ii}).

\newpage
\section{Details of D5 brane reflection matrices}\label{AppABs}
Invariance under the fermionic generators forces the functions appearing in (\ref{As})
to be
{\small
\ba
A_1(p) &\!\!= \!\!& {R}_0(p)
\nn\\
A_2(p) &\!\!= \!\!& {R}_0(p)\left(\frac{1}{3} + \frac{ \left((x^+)^2 - x_B x^-\right)\left(x_B(x^+)^2 +  x^-\right)}{(x^+)^2(x_B-x^-)(1+x_B x^-)}\right)
\nn\\
A_3(p) &\!\!= \!\!& - {R}_0(p) \frac{x_B x^++(x^-)^2}{x^-( x_B-x^-)}
\nn\\
A_4(p) &\!\!= \!\!& {R}_0(p) \frac{x_B(x^+)^2 + (x^-)^2(x_B (x^+)^2-2(x_B+x^+))
-x^- x^+ (x_B + x^+(3 x_B x^+ -2)) }{2x^-x^+(x_B-x^-)(1+x_B x^-)}
\nn\\
A_5(p) &\!\!= \!\!& {R}_0(p) \frac{ x_B x^--(x^+)^2}{x^-( x_B-x^-)}
\nn\\
A_6(p) &\!\!= \!\!& {R}_0(p)
\frac{x^-x^+\left(x_B + x^-(x_B x^-+4)\right)-x_B(x^-)^4 - x_B (x^+)^2 (2(x^-)^2 -1)}{2(x^-)^2(x_B-x^-)(1+x_B x^-)}
\nn\\
A_7(p) &\!\!= \!\!& -{R}_0(p) \frac{x^+(x_B+x^+)}{x^-(x_B-x^-)}
\nn\\
A_8(p) &\!\!= \!\!& {R}_0(p) \frac{ 2 x_B^2 (x^-)^3 + 2 (x^+)^3 -x^-x^+(x_B-x^+)(1-x_B x^-)}{(x^-)^2(x_B-x^-)(1+x_B x^-)}
\nn\\
A_9(p) &\!\!= \!\!& {R}_0(p) \frac{x_Bx^-x^+  +x_B(x^-)^3 x^+ +2(x^+)^2(x_B+x^+)
-(x^-)^2(x_B + x^+(3 x_B x^+ -2))}{2(x^-)^2(x_B-x^-)(1+x_B x^-)}
\nn\\
A_{10}(p) &\!\!= \!\!& {R}_0(p) \frac{ f x_B\left((x^+)^2 - (x^-)^2\right)^2}{2x^-x^+(x_B-x^-)(1+x_B x^-)\eta^2}
\nn\\
A_{11}(p) &\!\!= \!\!& {R}_0(p) \frac{ x_B(x^+ + x^-)^2 \eta^2}{2 f x^-x^+(x_B-x^-)(1+x_B x^-)}
\nn\\
A_{12}(p) &\!\!= \!\!& {R}_0(p) \frac{f x_B (x_B x^- -(x^+)^2)\left((x^+)^2 - (x^-)^2\right)}{\sqrt2 x^-x^+(x_B-x^-)(1+x_B x^-) \eta_B \eta }
\nn\\
A_{13}(p) &\!\!= \!\!& {R}_0(p) \frac{\sqrt2 \eta_B\eta  x_B (x_B x^- -(x^+)^2)(x^++ x^-)}{ f x^-x^+(x_B-x^-)(1+x_B x^-)}
\nn\\
A_{14}(p) &\!\!= \!\!& {R}_0(p) \frac{\eta  x_B (x_B (x^-)^2 -x^+)(x^+ + x^-)}{ 2 \sqrt2 \eta_B  x^-x^+(x_B-x^-)(1+x_B x^-)}
\nn\\
A_{15}(p) &\!\!= \!\!& {R}_0(p) \frac{\sqrt2 \eta_B  (x_B (x^-)^2 -x^+)\left((x^+)^2 - (x^-)^2\right)}{ \eta  (x^-)^2(x_B-x^-)(1+x_B x^-)}
\nn\\
A_{16}(p) &\!\!= \!\!& -{R}_0(p) \frac{ \eta_B  \eta  (x_B +x^+)(x^+ + x^-)}{2\sqrt2 f  x^-(x_B-x^-)(1+x_B x^-)}
\nn\\
A_{17}(p) &\!\!= \!\!& {R}_0(p) \frac{ f x_B (x_B +x^+)((x^+)^2 - (x^-)^2)}{\sqrt2 \eta_B  \eta  x^-(x_B-x^-)(1+x_B x^-)}
\nn\\
A_{18}(p) &\!\!= \!\!& {R}_0(p) \frac{\sqrt2 \eta x_B(x^++x^-)}{ \eta_B x^-(x_B-x^-)}
\nn\\
A_{19}(p) &\!\!= \!\!& {R}_0(p) \frac{\eta_B((x^+)^2-(x^-)^2)}{\sqrt2 \eta x^-(x_B-x^-)}
\ea
}
Similarly, for the reflection of the antisymmetric part of the bulk magnon, one finds the following coefficients
{\small
\ba
B_1(p) &\!\!= \!\!& \tilde {R}_0(p)
\nn\\
B_2(p) &\!\!= \!\!& \tilde {R}_0(p)  \left(\frac{1}{3} +
\frac{\left((x^-)^2 + x_B x^+\right)\left(x_B(x^-)^2 -  x^+\right)}{(x^-)^2(x_B + x^+)(1-x_B x^+)}\right)
\nn\\
B_3(p) &\!\!= \!\!& \tilde {R}_0(p) \frac{(x^+)^2 - x_B x^-}{ x^+(x_B+x^+)}
\nn\\
B_4(p) &\!\!= \!\!&  \tilde {R}_0(p)   \frac{x_B(x^-)^2 + (x^+)^2(x_B (x^-)^2-2(x_B-x^-))
-x^- x^+ (x_B + x^-(3 x_B x^- +2)) }{2x^-x^+(x_B+x^+)(1-x_B x^+)}
\nn\\
B_5(p) &\!\!= \!\!&  \tilde {R}_0(p) \frac{(x^-)^2 + x_B x^+}{x^+(x_B + x^+)}
\nn\\
B_6(p) &\!\!= \!\!&   \tilde {R}_0(p)
\frac{x^-x^+\left(x_B + x^+(x_B x^+ - 4)\right)-x_B(x^+)^4 - x_B (x^-)^2 (2(x^+)^2 -1 )}{2(x^+)^2(x_B+x^+)(1-x_B x^+)}
\nn\\
B_7(p) &\!\!= \!\!& -\tilde {R}_0(p) \frac{x^-(x_B-x^-)}{x^+(x_B+x^+)}
\nn\\
B_8(p) &\!\!= \!\!&  \tilde {R}_0(p) \frac{2 (x^-)^3 + 2 x_B^2 (x^+)^3+x^-x^+(x_B+x^-)(1+x_B x^-x^+)}{(x^+)^2(x_B+x^+)(1-x_B x^+)}
\nn\\
B_9(p) &\!\!= \!\!&     \tilde {R}_0(p)    \frac{x_Bx^-x^+ + x_B(x^+)^3 x^- +2(x^-)^2(x_B-x^-)
-(x^+)^2(x_B + x^-(3 x_B x^- +2))}{2(x^+)^2(x_B+x^+)(1-x_B x^+)}
\nn\\
B_{10}(p) &\!\!= \!\!& - \tilde {R}_0(p)  \frac{ x_B(x^+ + x^-)^2 \eta^2}{2 f x^-x^+(x_B+x^+)(1-x_B x^+)}
\nn\\
B_{11}(p) &\!\!= \!\!& - \tilde {R}_0(p)  \frac{ f x_B\left((x^-)^2 - (x^+)^2\right)^2}{2x^-x^+(x_B+x^+)(1-x_B x^+)\eta^2}
\nn\\
B_{12}(p) &\!\!= \!\!& \tilde {R}_0(p)  \frac{\eta_B\eta  x_B (x_B x^+ +(x^-)^2)(x^++ x^-)}{\sqrt2 f x^-x^+(x_B+x^+)(1-x_B x^+)}
\nn\\
B_{13}(p) &\!\!= \!\!& \tilde {R}_0(p)  \frac{\sqrt2 f x_B (x_B x^+ +(x^-)^2)\left((x^-)^2 - (x^+)^2\right)}{ x^-x^+(x_B+x^+)(1-x_B x^+) \eta_B \eta }
\nn\\
B_{14}(p) &\!\!= \!\!& \tilde {R}_0(p)   \frac{ \eta_B  (x_B (x^+)^2 +x^-)\left((x^-)^2 - (x^+)^2\right)}{2\sqrt2 \eta  (x^+)^2(x_B+x^+)(1-x_B x^+)}
\nn\\
B_{15}(p) &\!\!= \!\!& - \tilde {R}_0(p) \frac{ \sqrt2 \eta x_B (x_B (x^+)^2 +x^-)(x^+ + x^-)}{ \eta_B  x^-x^+(x_B+x^+)(1-x_B x^+)}
\nn\\
B_{16}(p) &\!\!= \!\!& \tilde {R}_0(p) \frac{f x_B(x_B -x^-)( (x^+)^2 -(x^-)^2)}{2 \sqrt{2} \eta_B \eta x^+ (x_B+x^+)(1-x_B x^+)}
\nn\\
B_{17}(p) &\!\!= \!\!& -\tilde {R}_0(p) \frac{ \eta_B  \eta  (x_B -x^-)(x^+ + x^-)}{\sqrt2 f  x^+(x_B+x^+)(1-x_B x^+)}
\nn\\
B_{18}(p) &\!\!= \!\!& \tilde {R}_0(p) \frac{\sqrt2 \eta_B((x^+)^2-(x^-)^2)}{\eta x^+(x_B+x^+)}
\nn\\
B_{19}(p) &\!\!= \!\!& \tilde {R}_0(p)  \frac{\eta x_B (x^+ +x^-)}{\sqrt2 \eta_B x^+(x_B+x^+)}
\ea
}


\end{document}